\documentclass{emulateapj}

\newcommand{\be}{  \begin{eqnarray} }
\newcommand{\ee}{  \end{eqnarray} }

\shorttitle{Disk Models of Black Hole X-ray Binaries}
\shortauthors{Davis et al.}
\begin{document}
\title{Relativistic Accretion Disk Models of High State Black Hole X-ray 
       Binary Spectra}
\author{Shane W. Davis\altaffilmark{1}, Omer M. Blaes\altaffilmark{1}, Ivan Hubeny\altaffilmark{2}, and Neal J. Turner\altaffilmark{3}}
\altaffiltext{1}{Department of Physics, University of California, Santa Barbara,
CA 93106}
\altaffiltext{2}{Steward Observatory and Department of Astronomy, University of Arizona, Tucson, AZ 85721}
\altaffiltext{3}{Jet Propulsion Laboratory, MS 169-506, California
Institute of Technology, Pasadena, CA 91109}
\begin{abstract}
We present calculations of non-LTE, relativistic
accretion disk models applicable to the high/soft state of black hole X-ray 
binaries. We include the effects of thermal Comptonization and bound-free and 
free-free opacities of all abundant ion species. Taking into account the
relativistic propagation of photons from the local disk surface to an observer
at infinity, we present spectra calculated for a variety of accretion rates,
black hole spin parameters, disk inclinations, and stress prescriptions.
We also consider nonzero inner torques on the disk, and explore different vertical
dissipation profiles, including some which are motivated by recent radiation MHD
simulations of magnetorotational turbulence.  Bound-free metal opacity generally 
produces significantly less 
spectral hardening than previous models which only considered Compton 
scattering and free-free opacity.  It also tends to keep the effective
photosphere near the surface, resulting in spectra which are remarkably
independent of the stress prescription and vertical dissipation profile,
provided little dissipation occurs above the effective photosphere.
This robustness is due to the fact that the disk scale height is largely
independent of the stress and dissipation when the disk is radiation
pressure supported and electron scattering dominated throughout its interior.
Our model
spectra may be affected by continuing theoretical uncertainties that we do not
account for, and we discuss these uncertainties.  We provide
detailed comparisons between our models
and the widely used multicolor disk model.  Frequency dependent
discrepancies exist that may affect the parameters of other spectral components
when this simpler disk model is used to fit modern X-ray data.  For a given
source, our models predict that the luminosity in the high/soft state should
approximately scale with the fourth power of the empirically inferred
maximum temperature, but with a slight hardening at high luminosities.
This is in good agreement with observations.
\end{abstract}


\section{Introduction}

The high state spectral energy distribution (SED) of black hole X-ray binaries
is dominated by a soft thermal component generally believed to be emission from
an optically thick accretion flow. 
It is often  hypothesized that the geometry of this flow
is a radiatively efficient, geometrically thin disk 
(Shakura \& Sunyaev 1973, hereafter SS73) or at higher luminosities a less
radiatively efficient slim disk 
(Abramowicz et al. 1998).  Most sophisticated spectral modeling
of black hole accretion disks has focused on the case of
supermassive black holes. However, models that compute
vertical structure and radiative transfer
have been constructed for Galactic black hole candidates assuming thin
disk (Shimura \& Takahara 1995, hereafter ST95; Merloni, Fabian \& Ross 2000) 
and slim disk (Wang et al. 1999) relations.

ST95 argue that a simple diluted blackbody provides 
adequate fits to the local specific flux $F_{\nu}$ for regions of the disk 
which produce most of the X-ray photons.  They conclude this for disks with 
sufficiently high luminosity over a range of 
luminosities, black hole mass, and black hole spin.  The local specific flux
is then given by
\be
F_{\nu}=\frac{\pi}{f_{\rm col}^4} B_{\nu}(f_{\rm col} T_{\rm eff})
\ee
where $B_{\nu}$ is the Planck function, $T_{\rm eff}$ is the effective 
temperature
of the emitting surface, and $f_{\rm col}$ is the color correction. 
They find that in many cases most of the disk spectrum is well 
approximated by a multi-temperature diluted blackbody with a single value 
of $f_{\rm col}$. Furthermore, the value of $f_{\rm col}$ is not
a strong function of the disk parameters.  They attribute these results
to saturated Comptonization playing a dominant role and producing
Wien-like spectral profiles. The relative constancy of 
$f_{\rm col}$ in these model annuli has been used to justify
fitting high 
state SED's with the multicolor blackbody model ({\it diskbb} in 
XSPEC)
of Mitsuda et al. (1984). Although this model assumes a simple profile of
effective temperature with radius, $T_{\rm eff}\propto R^{-3/4}$, and does
not account for relativistic transfer effects, it has been used to 
infer inner radii and black hole spins for disks believed to extend deep within
the gravitational potential of the black hole.

Merloni et al. (2000) caution against assuming a nearly constant value
for $f_{\rm col}$.  They
construct several constant density disk models with a range of accretion 
rates. They assume some fraction of the released gravitational energy 
dissipates directly into an optically thin corona above the disk (Svensson
\& Zdziarski 1994), but they do not account for the irradiation
of the disk surface by the coronal emission. They fit the resulting SED's 
with the {\it diskbb} model, apply relativistic corrections to the best
fit parameters, and use the results to infer an inner radius 
for the disk.  They find that the value of $f_{\rm col}$ needed to obtain 
the correct radius varies significantly from model to model.

Gierli\'nski \& Done (2004; hereafter GD04) investigate temperature and 
luminosity
evolution in a sample of black hole binaries observed in the high state
with the {\it Rossi
X-ray Timing Explorer} ({\it RXTE}).  They fit their X-ray spectra with
a model that includes a {\it diskbb} component and accounts for Comptonized
emission from a corona, focusing on observations in which the Comptonized
emission is $<15$\% of the bolometric luminosity.  They find that many of
the well observed sources are consistent with weak evolution of $f_{\rm col}$
in which $f_{\rm col}$ increases with higher luminosity or temperature.
Furthermore, they tentatively conclude that such evolution is inconsistent
with a modified stress prescription in which the accretion stress scales
with only the gas pressure and not the total pressure.

To our knowledge, all the detailed theoretical models of thermal disk 
spectra of black hole
X-ray binaries have so far neglected the effects of bound-free
opacity of metal ions.  (This is in marked contrast to models of
reflection spectra in these systems.)  Bound free opacity can
often dominate free-free opacity at observed X-ray energies,
reducing the relative importance of electron scattering and
therefore altering the expected color correction factors.  This
may therefore affect the fits to high/soft state spectra and
their interpretation.  Bound-free opacity also imparts
absorption/emission edges in the spectrum which are potentially
observable, particularly with the high throughput spectra being
produced by modern X-ray observatories.

In addition,
significant theoretical advances have been made recently in
understanding the nature and vertical distribution of turbulent
dissipation (Turner 2004), as well as highlighting the possibility
of significant external torques on the disk at the innermost
stable circular orbit (e.g. Gammie 1999; Krolik 1999; Hawley
\& Krolik 2002).  These results imply substantial modifications
of the basic assumptions underlying the disk models upon which
all spectral calculations have been based thus far.

In this paper, we construct fully relativistic accretion disk
model atmospheres and examine the integrated SED's from these
disks.  We fully account for bound-free opacity by incorporating
non-LTE rate equations for the ground-state level populations of 
all abundant metal ions.  We focus on understanding
the sensitivity of these SED's to the underlying assumptions in our
models.  We compare our results with previous theoretical investigations and 
recent observations.  In section \ref{method} we review the method used to
construct our models.
We describe our standard models in detail in section \ref{std}.  We 
report our results from altering the stress prescription in section 
\ref{stress}, adding an inner torque in section \ref{torque}, and modifying 
the dissipation profile in section \ref{dissip}.  In section \ref{lvst} we 
compare our model SED's with the work of GD04.  We provide further discussion 
of all these results and a summary of our conclusions in sections 
\ref{discus} and \ref{conc}.  Readers primarily concerned with the
observational implications of our models may wish to skip directly to sections
\ref{lvst}, \ref{discus}, and \ref{conc}.

\section{Method}
 \label{method}
We construct models using the methods described in a series of papers
(Hubeny \& Hubeny 1997, 1998, hereafter HH98; Hubeny et al. 2000, 
Hubeny et al. 2001,
hereafter HBKA). We construct thin disk models by solving the fully
relativistic one zone disk structure equations in the Kerr metric 
(Novikov \& Thorne 1973, hereafter NT73; Page \& Thorne 1974; Riffert 
\& Herold 1985). Next, we compute the vertical structure and local spectra
for annuli evenly spaced in log $r$ where $r=R/R_{\rm g}$, $R$ is the 
Boyer-Linquist radial coordinate, and $R_{\rm g}=GM/c^2$ is the gravitational
radius. We compute ten annuli per decade in radius and assume the disk 
extends from the radius of 
marginal stability $r_{\rm in}=r_{\rm ms}$ to $r_{\rm out} \approx 1000$.  
Emission from radii larger than $r_{\rm out}$ is assumed to be blackbody and 
contributes little to the integrated flux at photon energies $\gtrsim 0.1$ 
keV.

At each annulus, we use the TLUSTY stellar atmospheres code 
(Hubeny \& Lanz 1995) to simultaneously
solve the equations for the vertical structure and angle dependent radiative 
transfer. At X-ray temperatures, metal opacities are important, and we
incorporate fully non-LTE ground state level populations for all ions of 
H, He, C, N, O, Ne, Mg, Si, S, Ar, Ca, Fe and Ni assuming solar abundances.
Bound-free opacity due to each ion is included but
bound-bound transitions are neglected.  We account for Comptonization with an
angle-averaged, Kompaneets treatment of the electron scattering source
function.  Finally, we calculate the integrated disk spectrum seen by an 
observer at infinity by using a fully general-relativistic transfer
function (Agol 1997).

We make several approximations which may be violated in nature. 
The disk models are time steady and azimuthally symmetric. All heat
is transported vertically by radiative energy flux. There is no mass 
loss from the disk, so the accretion rate is independent 
of radius.  All radiative flux is assumed to originate within the disk and its
atmosphere; irradiation of the disk surface due to an external X-ray source is
neglected.  Irradiation of the disk surface by returning radiation
(Cunningham 1976)  is also ignored.  The atmosphere calculations are all 
one-dimensional despite evidence that radiation pressure supported
accretion disks have significant inhomogeneity (Turner et al. 2003,
Turner 2004).

For our base model (hereafter the standard model) we make additional 
assumptions which we later
relax. First, there is no torque on the inner edge of the
disk.  Second, the energy dissipation rate per unit volume 
$\epsilon = \bf{\nabla} \cdot \bf{F}$ is locally proportional to the 
density $\rho$.
Third, we relate the vertically averaged stress, $\bar{\tau}_{r\phi}$, to
the vertically averaged total pressure, $\bar{P}$ using the prescription (SS73)
\be
\label{eq:stress}
\bar{\tau}_{r\phi}=\alpha \bar{P}.
\ee

This standard model is parameterized by only four quantities: black hole mass
$M$, spin parameter $a$, the stress parameter $\alpha$, and accretion rate 
(SS73; NT73).
The accretion rate is often expressed relative to the Eddington 
accretion rate as
$\dot m = \dot M c^2/L_{\rm Edd}$ where $L_{\rm Edd}=1.5 \times 10^{38} 
(M/M_{\odot}) 
\rm \, erg \, s^{-1}$ is the Eddington luminosity for completely ionized H.  We
prefer to use the quantity $l \equiv L/L_{\rm Edd}=\eta \, \dot m$ 
as it relates more directly to the observed luminosity.
Here $L$ is the bolometric luminosity of the disk and $\eta(M,a)$ is the 
fraction of the energy at infinity which is radiated before crossing 
$r_{\rm ms}$.
The mass is fixed at $10 M_{\odot}$ for all models considered here.
 
The atmospheres of the individual annuli are entirely determined by specifying 
the midplane column mass, the radiative flux, the local gravity, the 
composition and the vertical dissipation profile $\epsilon(z)$. The midplane column
mass $m_{0}$ is equal to half the surface
density of the disk.  The flux is parameterized by the effective temperature,
$T_{\rm eff}$, of the corresponding blackbody emitter.  The vertical component of 
the gravitational acceleration $g=Q \, z$ is proportional to the distance 
$z$ above the midplane and the constant of proportionality, $Q$, is a function 
only of $R$.  These quantities
($m_{0}$, $T_{\rm eff}$, $Q$) are uniquely determined at a radius $r$ in a one-zone 
model by a given set of disk parameters ($M$, $l$, $\alpha$, $a$). 
As noted above, we consider $\epsilon(z) \propto \rho(z)$ for
the standard models, but explore other prescriptions in section \ref{dissip}.

Except for the method of calculating $m_0$, the standard models are
equivalent to the supermassive black hole
models described in HBKA and the reader is referred there for further details.
HBKA solved a one-zone, algebraic relation for $m_0$ in terms of $M$, $a$, 
$\alpha$, $l$, and $r$. This $m_{0}$ is then used
throughout the stellar atmospheres calculation. Due to the simplifying
assumptions of the one-zone model, there is no guarantee that equation 
\ref{eq:stress} will hold for the converged solution.  The vertically integrated
pressure $\bar{P}$ can be calculated from the converged non-LTE model so the 
consistency  of this prescription can be checked after the fact using equation 
\ref{eq:stress}.
The one-zone calculation provides a suprisingly accurate approximation,
but order unity discrepancies do typically arise between the final $\alpha$ 
calculated from the converged non-LTE model and the input $\alpha$.. 
 
We alter this prescription by using the initial $m_0$ 
to construct the LTE-grey model (Hubeny 1990) from which we calculate 
an $\alpha'$ with equation \ref{eq:stress}.  If $\alpha'$ is within 10\% of 
$\alpha$ we continue the 
full atmosphere calculation using the initial $m_0$.  Otherwise, we iterate 
by choosing a new $m_{0}$, recomputing the LTE-grey model and $\alpha'$ 
until this condition is satisfied. We use Newton-Raphson to calculate
the change in $m_0$ for the next iteration: 
$\delta m_0=dm/d \alpha (\alpha-\alpha')$ with the derivative 
evaluated analytically assuming the one zone scaling holds.
The resulting LTE-grey model is then used as the starting point for
constructing the full non-LTE atmosphere. For most annuli, this method rapidly
converges and provides a final $\alpha$ consistent (within $\sim 10\%$) with 
the input value.
However, in some of the hottest annuli of the torqued disk model,
the method fails to converge due to inaccuracies in the LTE-grey model. In 
this case, we default to the one-zone calculation
for {\it all annuli} in the disk, but in all other cases we utilize the 
iterative procedure.

\section{The Standard Model}
\label{std}
\subsection{The Local Spectrum \label{local}}

The disks in X-ray binaries with a black hole accretor can be much hotter
than a disk in a cataclysmic variable or quasar.
The gas temperatures at the surface of these disks are sufficiently
high ($\sim 10^6$K - $10^7$K) that electron scattering 
dominates the opacity at typical photon energies and the emitted spectrum 
may deviate significantly from that of a blackbody (SS73).

We define the effective optical depth as 
\be
\tau^{\rm eff}_{\nu} \equiv \int^m_0 \left[3 \kappa^{\rm th}_{\nu}(\kappa^{\rm es}_{\nu}+
	\kappa^{\rm th}_{\nu})\right]^{1/2} \, dm'
\ee
where $\kappa^{\rm th}_{\nu}=\kappa^{\rm ff}_{\nu}+\kappa^{\rm bf}_{\nu}$ is sum of the
total bound-free and free-free absorption opacity, $\kappa^{\rm es}_{\nu}$ is
the electron scattering opacity, and $m$ is column mass measured from the
surface. A closely related quantity is the depth of formation
\be
\tau^{\ast}_{\nu} = \int^{m^{\ast}_{\nu}}_0 (\kappa^{\rm es}_{\nu}+
	\kappa^{\rm th}_{\nu}) \, dm'
\label{eq:taustar}
\ee
where $m^{\ast}_{\nu}$ is the {\it frequency dependent} column mass at which 
$\tau^{\rm eff}_{\nu}=1$.
All of the disk models considered in this paper are effectively thick 
(i.e. $m^{\ast}_{\nu} < m_{0})$ at each annulus for frequencies of interest.
In an electron scattering dominated atmosphere 
$\kappa^{\rm th}_{\nu} \ll \kappa^{\rm es}_{\nu}$ for typical photon energies
implying that the photon destruction probability,
$\epsilon_{\nu}=\kappa_{\nu}^{\rm th}/(\kappa_{\nu}^{\rm th}+\kappa_{\nu}^{\rm es}) \ll 1$ and $\tau^{\ast}_{\nu} \gg 1$.  Due to the temperature gradients
in the atmosphere, the temperature at the depth of
formation $T_{\nu}^{\ast}$ is generally greater than $T_{\rm eff}$.
In the absence of Comptonization, the frequency dependence of $\epsilon_{\nu}$
and $T^{\ast}_{\nu}$ alter the spectrum and  produce a modified blackbody.

Due to the potentially large number of scatterings ($n_{\rm es} \approx 
\tau^{\ast 2}_{\nu} \gg 1$), Comptonization may also be important. To
account for its effect on our model spectra we
{\it always} solve the radiative transfer equation including the
Compton scattering source term. However, we still
wish to have an indicator of the impact of  electron scattering
on the SED. For a homogeneous electron distribution, the $y$-parameter 
provides a simple
means of characterizing the importance of Compton scattering.  To gain
similar insight, we generalize this prescription by 
defining a frequency dependent $y$-parameter (HBKA)

\be
y^{\ast}_{\nu} \equiv \frac{4 k_{\rm B} T^{\ast}_{\nu}}{m_{\rm e} c^2} {\rm max} [\tau^{\ast 2}_{\nu},\tau^{\ast}_{\nu}]
\label{eq:yeff}
\ee
where $k_{\rm B}$ is Boltzman's constant, $T^{\ast}_{\nu}$ is the temperature 
at the depth of formation, $m_{\rm e}$ is the electron rest mass, and $c$ is 
the speed of light.

\begin{figure}
\plotone{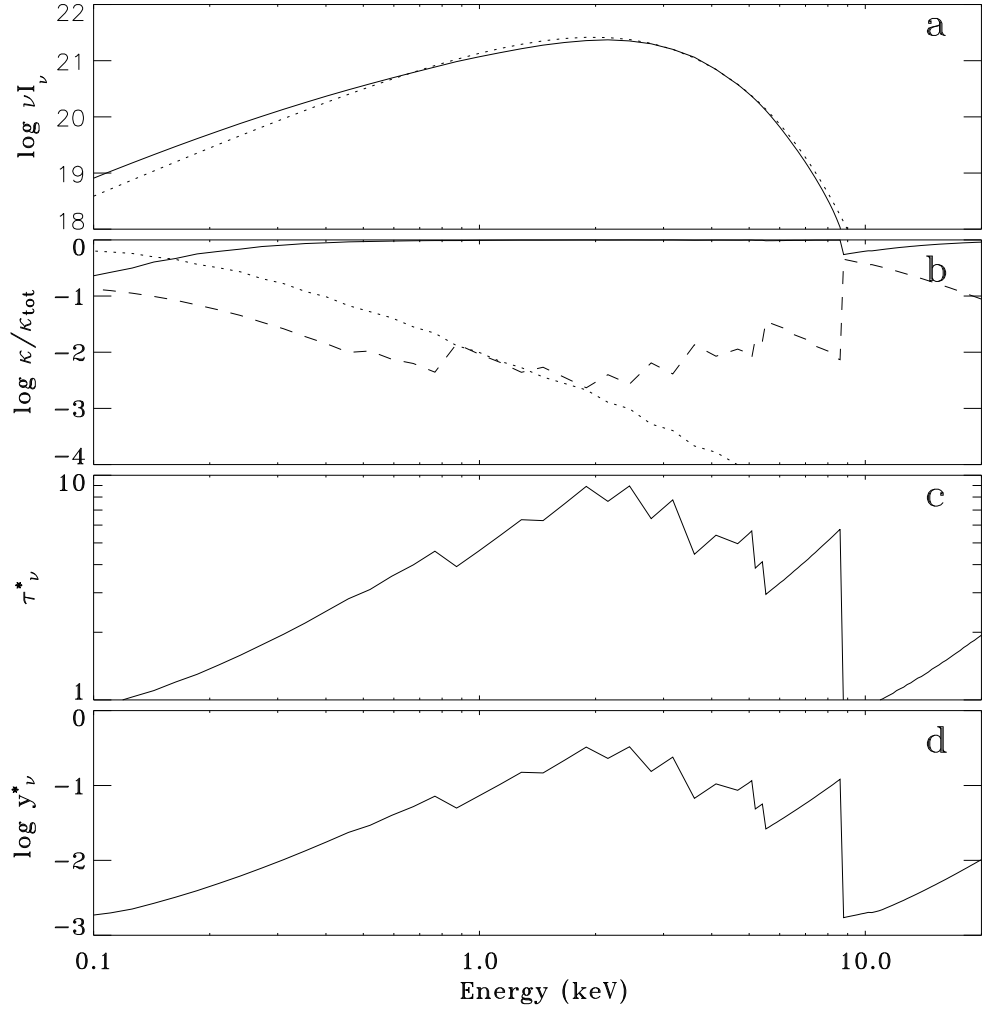}
\caption{
Specific intensity and other frequency dependent parameters evaluated
in the local frame at $r=12.6$ for a disk with $\alpha=0.1$, $l=0.1$, and
$a=0$.  (a) We compare the specific intensity viewed from $55^{\circ}$ 
(solid curve) with a diluted blackbody at the same $T_{\rm eff}$ and 
$f_{\rm col}=1.56$ (dotted curve).
The units of the ordinate are ergs s$^{-1}$ cm$^{-2}$ ster$^{-1}.$
(b) We show the fraction
of the total opacity provided by electron scattering (solid curve), free-free
absorption (dotted curve), and bound-free absorption (dashed curve)
evaluated at $m^{\ast}_{\nu}$. (c) The depth of formation (eq. 
\ref{eq:taustar}) is plotted.  (d) The frequency dependent $y$-parameter
(eq. \ref{eq:yeff}) is plotted.
\label{fig:depth}}
\end{figure}

In Figure \ref{fig:depth}, we plot the locally emitted SED from a single 
annulus with the frequency dependent quantities defined above.  The model 
atmosphere was calculated for an annulus at $r=12.6$ in a disk with $l=0.1$
and $\alpha=0.1$ accreting onto a Schwarzschild black hole.  This 
is near where the flux (and thus the effective temperature) of the disk peaks.
In the top panel (a) we compare the intensity at an inclination $i$ of 
$55^{\circ}$ to the vertical with that emitted by a diluted blackbody
at the same $T_{\rm eff}$ with $f_{\rm col}=1.56$. 
The SED is a modified blackbody with a spectral break at the 
He-like Fe photoionization edge near 8.8 keV.  Below this edge and above
$\sim 2$ keV, the spectrum is well
approximated by the diluted blackbody.  In (b) we plot the
relative contributions of electron scattering, free-free absorption,
and bound-free absorption to the total opacity. The
opacities are evaluated at $m^{\ast}_{\nu}$. As expected,
electron scattering dominates except at low energies and just blueward of
the He-like Fe edge.  Bound-free exceeds free-free absorption above $\sim$1 
keV and therefore should not be neglected. Note that a decrease in
our frequency grid resolution below $\sim 6$ keV contributes to the
breadth of the photoionization edges in that range. Panel (c) shows
how $\tau^{\ast}_{\nu}$ depends strongly on the presence of metals.  As
photon energy increases, the bound-free opacity overtakes the free-free 
opacity, preventing the depth of formation from exceeding 10.  If we had 
neglected metal opacities, the depth of formation would continue to grow as the
ratio of free-free to scattering opacity declined. In 
that case, the photons reaching an observer are created deeper in the 
atmosphere where the temperatures
and densities are higher.  A direct consequence is seen in (d)
where $y^{\ast}_{\nu} < 1$ for all frequencies of interest.
Compton scattering may contribute to the broadening of photoionization edges, 
but it is {\it not} significantly altering the continuum.  If we had neglected 
metals, the depth of formation would have been larger, the number of 
scatterings would have been higher, $y^{\ast}_{\nu}$ would have been greater 
than unity for $h \nu \gtrsim 2$ keV, and Comptonization would have had a 
larger effect. 

Despite the limited role of Comptonization, the specific intensity is still
reasonably well approximated by the diluted blackbody spectrum. A modified 
blackbody caused by
dominant electron scattering opacity, strong temperature gradients, or both is 
generally not well characterized by a single temperature so this result
is somewhat surprising.  It seems to be due to the relative
constancy of $\tau^{\ast}_{\nu}$ over the range of photon energies near
the peak.  This means that most of the photons are being created by gas at a
narrow range of temperatures and that the photon destruction probability
is only weakly dependent on frequency.  Both of these effects lead to a more
blackbody like spectral profile but $T^{\ast}_{\nu}$ is still larger than 
$T_{\rm eff}$.  In this case, $T^{\ast}_{\nu} \approx 1.5-1.6 T_{\rm eff}$
near the peak consistent with the $f_{\rm col}=1.56$ used for the plot. If
we had neglected bound-free metal opacity and Comptonization, the absorption 
opacity would be more frequency dependent and the diluted blackbody would be 
a poorer match to the emission.

\begin{figure}
\plotone{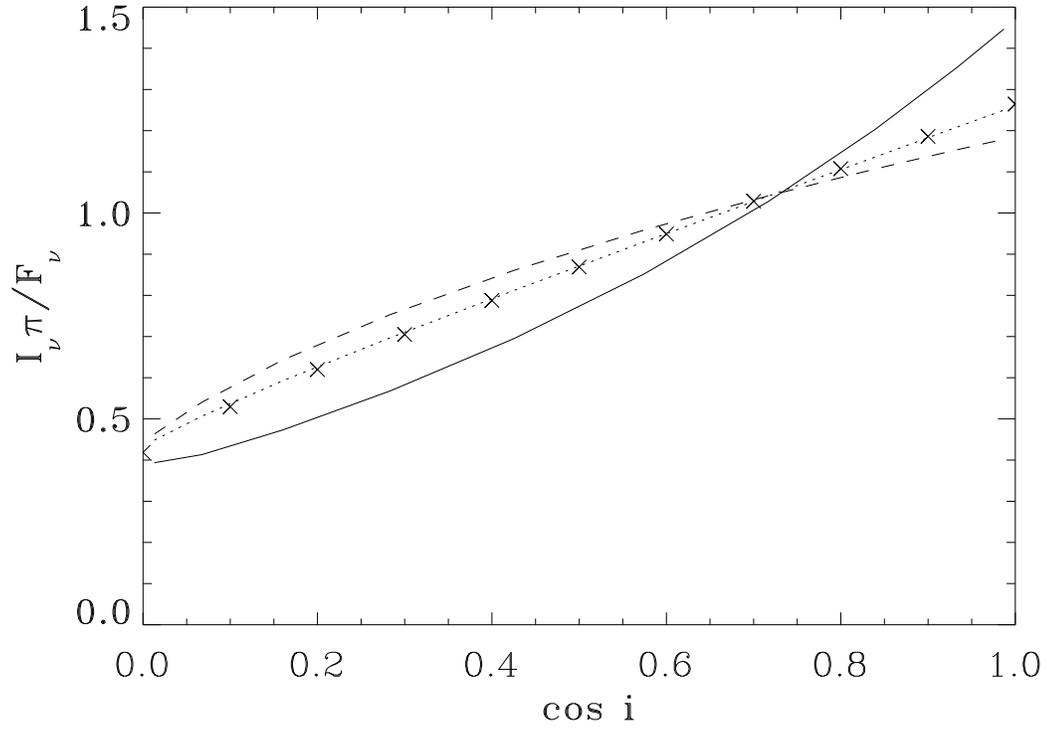}
\caption{Normalized specific intensity of the emission at at $r=12.6$ in 
an $l=0.1$, $\alpha=0.1$, and $a=0$ disk model. The curves
are normalized by a factor of $F_{\nu}/\pi$ so that isotropic emission would
have an ordinate of unity. The curves are plotted for photon energies of
0.21, (dashed) 2.2. (dotted), and 8.4 keV (solid).  For comparison we
plot the limb darkening law (x's) for a semi-infinite Thompson
scattering atmosphere assuming a Rayleigh phase function (Chandrasekhar 1960).
\label{fig:limbd}}
\end{figure}

Electron scattering and temperature gradients also affect the angular 
dependence of the radiation field. As shown in Figure \ref{fig:limbd},
the spectra are significantly limb darkened at frequencies above 0.1 keV
and the degree of limb darkening is frequency dependent. The curves in
Figure \ref{fig:limbd} show the normalized specific intensity for emission
at $r=12.6$ from an $l=0.1$, $\alpha=0.1$, and $a=0$ disk model.  The curves
are normalized by the specific flux so that isotropic emission would
have an ordinate of unity.  (The reason we chose $i=55^\circ$ in Figure 1 is
that the limb darkened emission matches isotropic emission near this viewing
angle.) The curves are plotted for photon energies of
0.21, 2.2., and 8.4 keV.  In this case, the emission near the peak (2.2 keV)
is well approximated by the results for a semi-infinite Thompson scattering
atmosphere assuming a Rayleigh phase function (Chandrasekhar 1960).  It is 
less limb darkened at lower frequencies and more limb darkened at higher 
frequencies due to the temperature gradient in the atmosphere. 

\subsection{Disk Integrated Spectra \label{lumin}}

We calculate four standard model disks with $\alpha=0.1$ accreting onto a 
Schwarzschild black hole with $l$=0.01, 0.03, 0.1, and 0.3.  Although larger 
values of $l$ are astrophysically interesting, we consider only models with 
$l \le 0.3$. For $l \sim 1$, radial advection begins to become
important and the assumption that all locally dissipated accretion power 
is radiated locally ceases to be accurate. In this regime, a slim disk model
should be used in place of the thin disk model which is no longer 
self-consistent.

\begin{figure}
\plotone{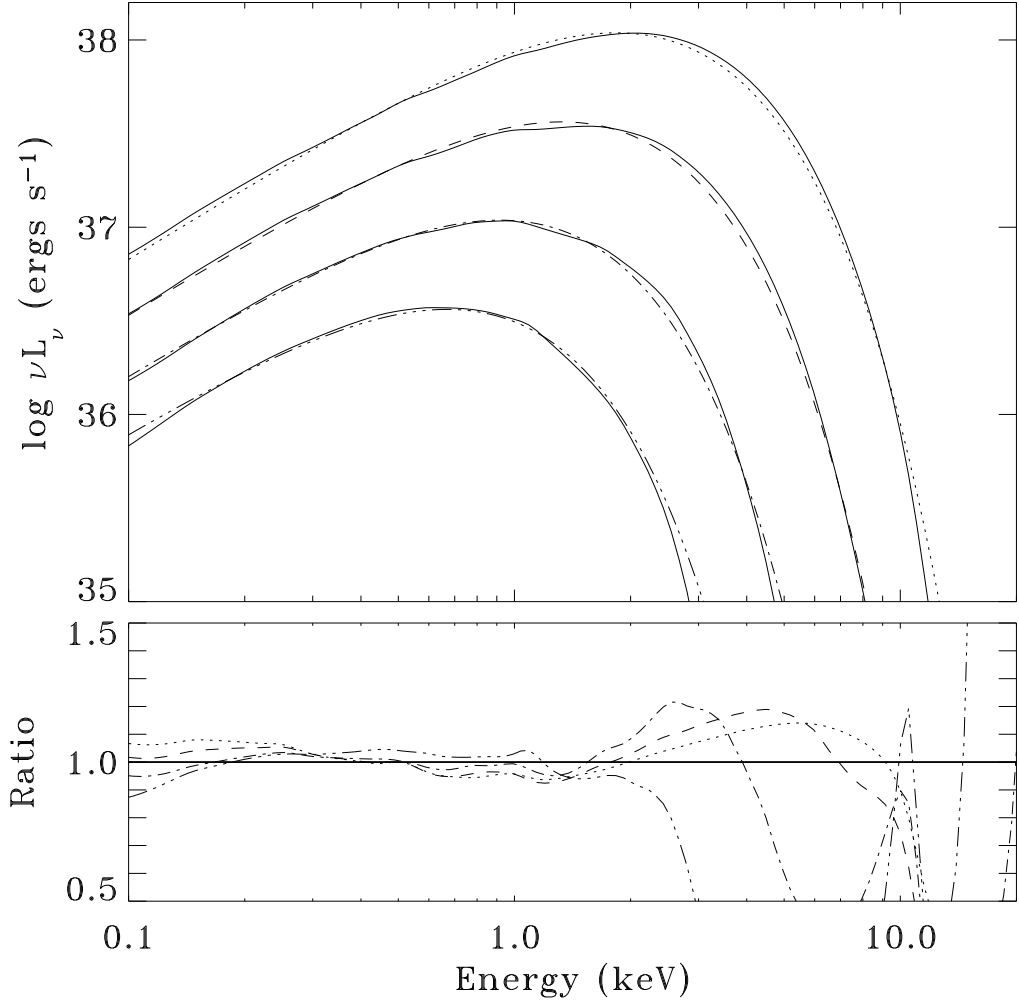}
\caption{Integrated disk SED's of four values of $l$ 
(solid) viewed at $i=70^{\circ}$ by an observer at 
infinity.  At each $l$, we also plot the best-fit, fully-relativistic disk 
model spectra
in which the local flux is assumed to be an isotropic diluted blackbody.
Each model has
$\alpha=0.1$ and $a=0$ and from lower left to upper right the curves correspond
to $l=0.01$, 0.03, 0.1, and 0.3. with best fit $f_{\rm col}=1.4$ 
(triple dot-dashed), 1.46 (dot dashed), 1.56 (dashed), and 1.64 (dotted)
respectively.  The diluted blackbody spectra are plotted at 81\%
of their intrinsic luminosity to account for limb darkening in the non-LTE
atmosphere models. The lower panel shows the ratio of the non-LTE model
spectra to the diluted blackbody spectra.}
\label{fig:lspec} 
\end{figure}

In Figure \ref{fig:lspec} we plot the standard model SED's (solid curves) 
observed from infinity at $i=70^{\circ}$. At each value of $l$, we also plot
best-fit, fully-relativistic spectra from Schwarzschild disks in which
the local SED at each annulus is assumed to be a diluted blackbody. These diluted
blackbody disks have the same surface brightness and utilize the same
transfer function as the standard model disks.  Therefore, they do not directly
match the {\it diskbb} model which assumes a simplified form for the 
radial dependence of the flux and neglects relativistic transfer effects.
The best fit values of $f_{\rm col}=1.4$, 1.46, 1.56, and 1.62 for the
$l=0.01$, 0.03, 0.1, and 0.3 models respectively.  Because of limb darkening,
the fully non-LTE atmosphere spectra have a lower apparent luminosity than
the isotropic diluted blackbodies at $i=70$.  To account for this, the 
diluted black body spectra are plotted at 81\% of their intrinsic 
luminosity.

We performed the fits shown in Figure 3 by creating XSPEC table models from diluted 
blackbody spectra and the fully non-LTE spectrum at the same $l$
and $i$.  For each $l$ and $i$, we use the non-LTE model SED to generate an 
artificial PHA data set using a diagonal response matrix with constant 
effective area.
This data set is then fit over the 0.1-10 keV range with the corresponding 
diluted blackbody table model which has two parameters: $f_{\rm col}$ and
the normalization. We account for the effects of limb darkening in the non-LTE 
atmospheres by fitting for the normalization of the diluted blackbody model
and find that the best fit models deviate from the artificial spectra by less 
than 10-20\% except out in the high energy tail.  (In an observed source, this
tail would have low photon statistics.)  If we had not fit the 
normalization the deviations would be as large as 50\% at higher inclinations.

\begin{figure}
\plotone{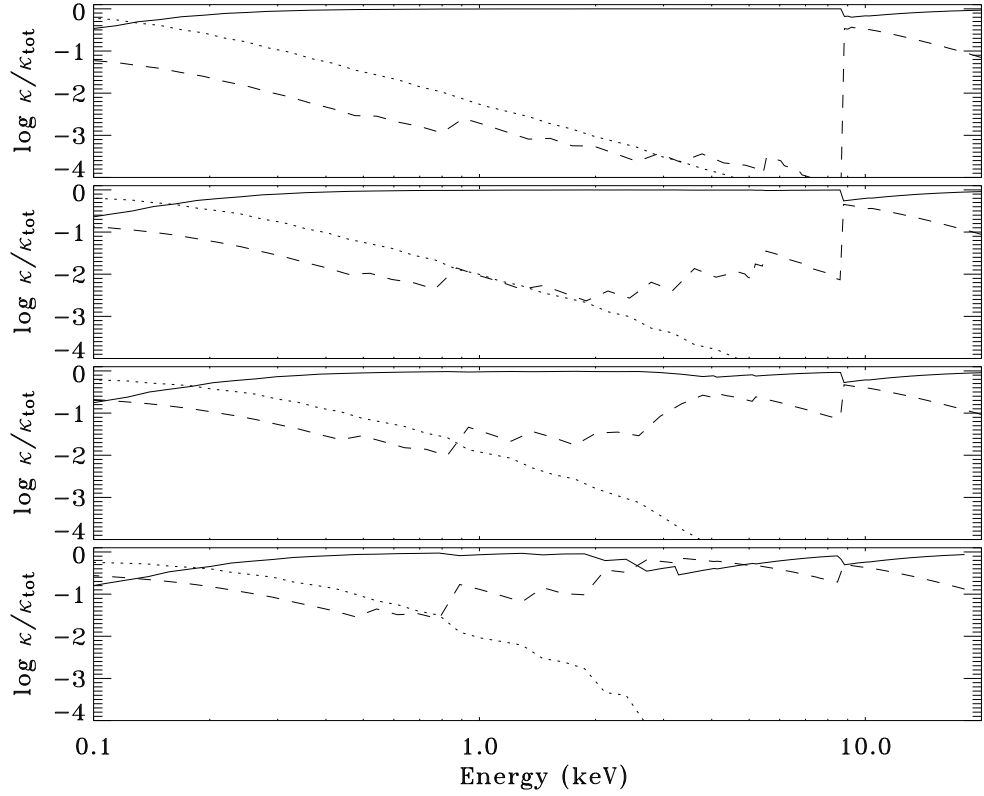}
\caption{
Fraction of the total opacity provided by electron scattering 
(solid), free-free absorption (dotted), and bound-free absorption 
(dashed) evaluated at $m^{\ast}_{\nu}$ and  $r=12.6$ in an 
$\alpha=0.1$ and $a=0$ disk. From top to 
bottom, the panels correspond to $l$=0.3, 0.1, 0.03, and 0.01.
\label{fig:labs}}
\end{figure}

Because $a$ and $M$
are fixed, the emitting area of the disk is fixed. An increase in luminosity
can only result from an increase in the surface brightness of the disk surface.
For blackbody emission, $\nu_{\rm peak} 
\propto T_{\rm eff}(r_{\rm max}) \propto l^{1/4}$.
The same general trend is observed for our models, but the scaling
does not strictly hold due to deviations from isotropic blackbody emission
in the atmosphere calculations.
As $l$ increases, it can be seen that standard model SED's become 
increasingly hard relative to a blackbody disk at the same luminosity. 
That is, the solid curves with high (low) $l$ are better fit with 
diluted blackbodies with higher (lower) values of $f_{\rm col}$. This relative 
hardening is largely due to the increased dominance of electron scattering 
in the more luminous models. The trend can be seen in Figure \ref{fig:labs}
where we plot 
the relative contributions of the opacity mechanisms at $\tau^{\ast}_{\nu}$ in
the $r=12.6$ annuli.  As $l$ increases, the radiative flux increases and the 
metals become highly ionized.  The corresponding decrease in the  bound-free 
opacity at lower photon energies moves $\tau^{\ast}_{\nu}$ nearer the midplane
where the temperatures are larger.  This produces a modified blackbody
spectrum which is harder for higher values of $l$.  For $l=0.3$, only the 
highest ionization states of iron are sufficiently populated to provide 
significant bound-free opacity.  This produces a larger value for 
$\tau^{\ast}_{\nu}$ and $y^{\ast}_{\nu} \gtrsim 1$,
implying that Comptonization is saturated.

\begin{figure}
\plotone{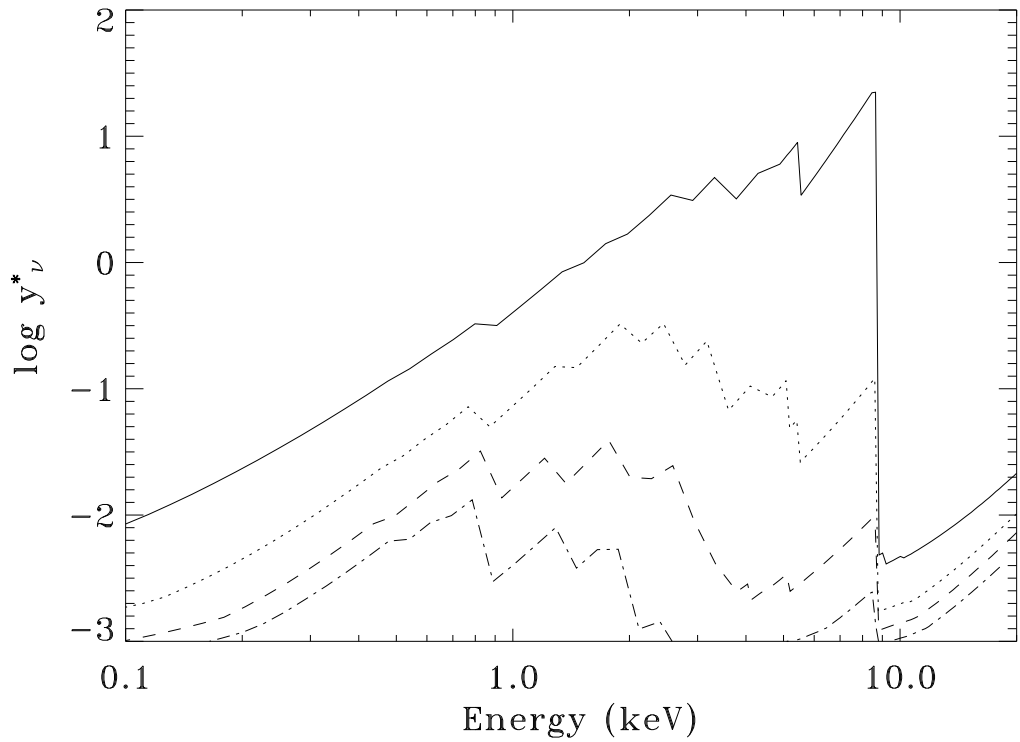}
\caption{Frequency dependent $y$-parameter, $y^{\ast}_{\nu}$, for models
with four different values of $l$ evaluated in the local frame of the disk
at r=12.6..
The curves correspond to $l$=0.3 (solid), 0.1 (dotted), 0.03
(dashed), and 0.01 (dot-dashed).
\label{fig:yeff}}
\end{figure}

ST95 attribute the effectiveness of the diluted blackbody approximation for
their higher luminosity models to the presence of saturated Comptonization 
producing Wien-like spectra in the inner annuli of their disks.
The impact of Compton scattering on our spectral calculations can be 
approximately estimated
from Figure \ref{fig:yeff} in which we plot $y^{\ast}_{\nu}$ in 
the $r=12.6$ annulus of each model. We find that $y^{\ast}_{\nu} \gtrsim 1$ 
only in the $l=0.3$ model (solid line). Nevertheless, the diluted blackbody 
spectra still deviate from the non-LTE model spectra by at most 10-20\% at 
typical photon energies above 0.1 keV.  As discussed in section \ref{local},
this is in part due to the relatively
weak frequency dependences of $T^{\ast}_{\nu}$ and $\epsilon_{\nu}$ near
the spectral peak. 
The agreement between the two types of spectra generally improves with 
increasing $l$. The effects of increasingly saturated Comptonization may
account for some of this trend. The reduction in bound-free opacity 
as the degree of ionization increases probably also plays a role since some
of the deviations are associated with spectral breaks at the photoionization
edges.

\begin{figure}
\plotone{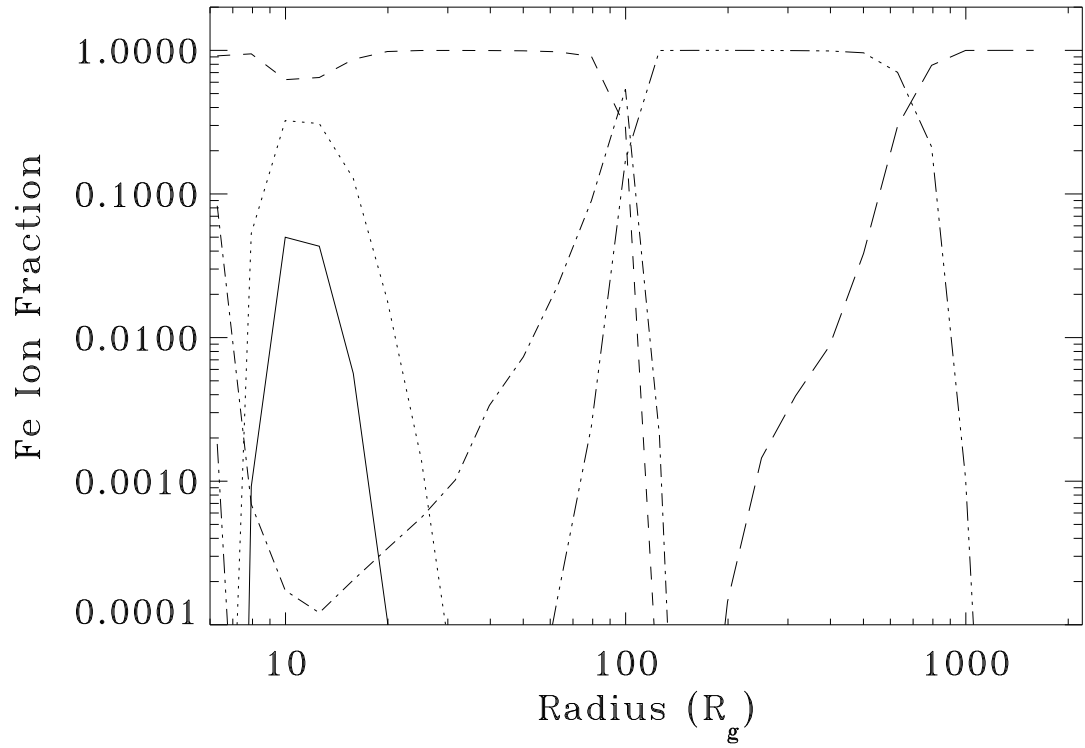}
\caption{
Ion fraction of iron at the Thompson photosphere as a function of radius for an 
$l=0.3$ and  $\alpha=0.1$ disk accreting onto a Schwarzschild black hole.
The curves correspond to Fe~XXVII (solid), Fe~XXVI (dotted), Fe~XXV (short
dashed), Fe~XXIV (dot-dashed), Fe~XVII - Fe~XXIII (triple dot-dashed), and 
Fe~I - Fe~XVI (long dashed).
\label{fig:ion}}
\end{figure}

In constructing these SED's we have neglected the irradiation of the disk surface
by an external corona.  Provided the local ionization parameter of the irradiating
coronal flux is sufficiently low, the reflection spectrum will be determined 
predominantly by the intrinsic ionization state of the disk.  For illustration,
we show the ionization fraction of iron at a Thompson depth of unity in Figure
\ref{fig:ion}, for the $l=0.3$ disk.  
Inside $r=100$ Fe~XXV is the dominant ion, with a small fraction of Fe~XXVI and
Fe~XXVII at $r \gtrsim 10$.  The drop in ionization state near $r \sim 6$ is
due to the no-torque inner boundary condition.  For this disk model, we expect the 
Fe K$\alpha$ line features in the reflection component to be dominated by the Fe~XXV
recombination line.  Resonant trapping limits fluorescent line emission for
Fe~XVII through  Fe~XXIII (Ross \& Fabian 1993) so any feature from more neutral
iron must come from $r \gtrsim 1000$.

\subsection{Dependence on Spin \label{spin}}

In an X-ray binary, the case of non-zero black hole spin seems particularly
plausible.  Here we consider the
``maximally spinning'' $a=0.998$ case (Thorne 1974) as an upper bound on the
disk spin with the expectation that angular momentum extraction mechanisms
likely limit real black holes to lower spin parameters.

Changing $a$ modifies the one-zone radial disk structure equations and the
location of $r_{\rm ms}$ (NT73).
Near the event horizon, the relativistic effects
on the photon geodesics are amplified, altering the SED observed from infinity.
At fixed $l$, the accretion rate drops due to the increased efficiency.
However, the innermost annuli are still hotter than in the Schwarzschild case
and the SED is expected to be harder modulo relativistic effects on the 
geodesics.

\begin{figure}
\plotone{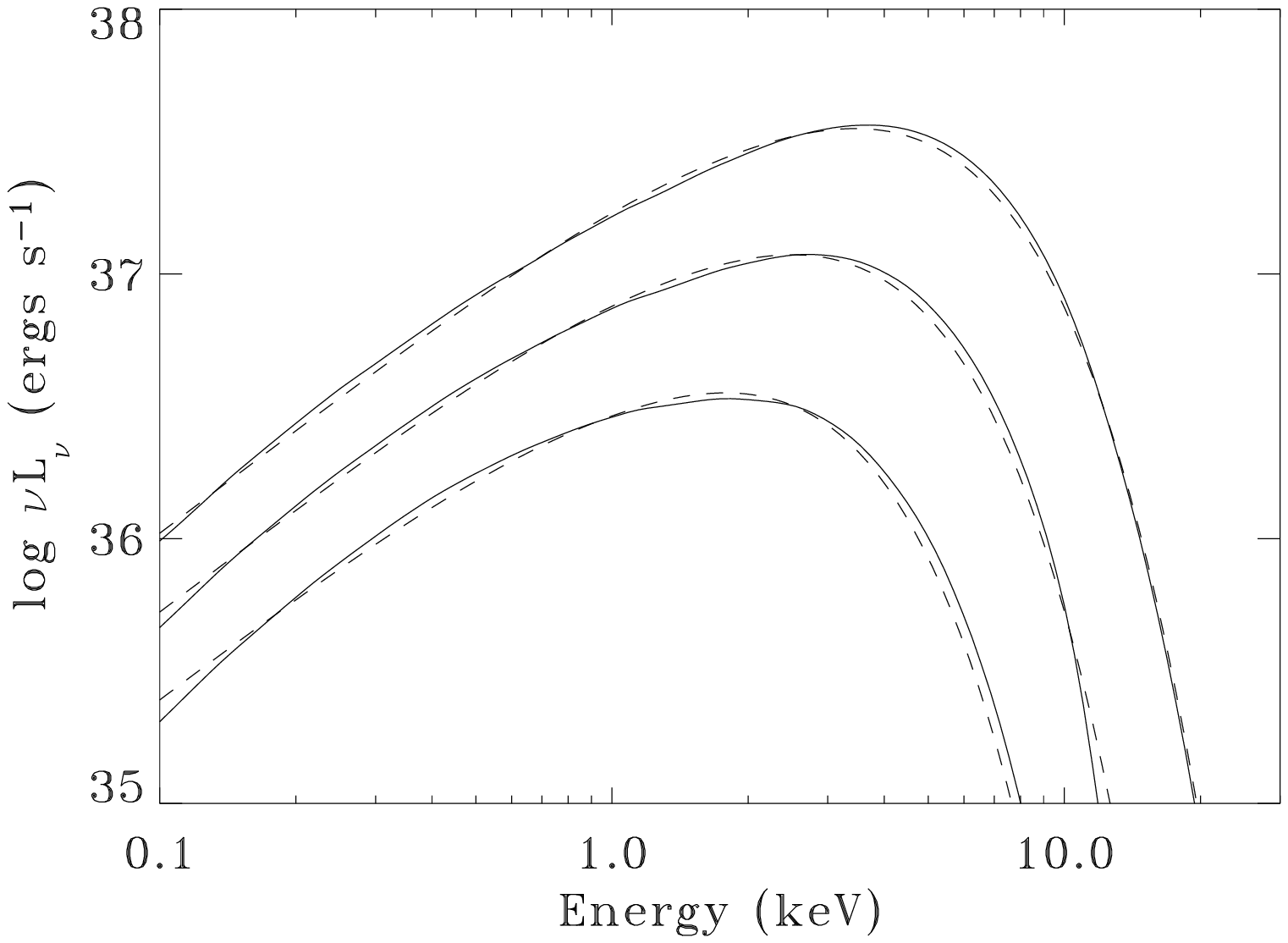}
\caption{
Integrated disk SED's for three values of $l$ (solid) observed from an 
$i=70^{\circ}$ by an observer at infinity.  At each $l$, we also plot the best-fit,
fully relativistic disk models in which the local flux is assumed to be an 
isotropic diluted blackbody (dashed). Each model has
$\alpha=0.1$ and $a=0.998$ and from lower left to upper right the curves 
correspond to $l=0.01$, 0.03, and 0.1 with best fit $f_{\rm col}=1.47$, 1.53, and 1.6
respectively. The diluted blackbody spectra are plotted at 95\%
of their intrinsic luminosity to account for limb darkening in the non-LTE
atmosphere models.
\label{fig:lkerr0.34}}
\end{figure}

We construct three maximally spinning Kerr models with $\alpha=0.1$ 
accreting at $l=0.01$, 0.03,
and 0.1.  Difficulties obtaining convergence in some of the inner annuli have
so far prevented us from completing an $l=0.3$ disk.
In Figure \ref{fig:lkerr0.34}, we plot the integrated SED's for these
models observed at
$i=70^{\circ}$ by an observer at infinity. At each value of $l$, we also 
plot the best-fit, fully relativistic spectra from maximally spinning disks 
in which the spectrum at each annulus is assumed to be a diluted blackbody. 
We fit them with the method described in section \ref{lumin}.
Comparison with Figure \ref{fig:lspec}
shows that both the full atmosphere models and the diluted blackbody curves
are harder than their Schwarzschild counterparts at the same $l$. Increasing $a$ modifies
the one-zone radial disk structure equations, yielding larger fluxes and correspondingly 
higher $T_{\rm eff}$ in the innermost annuli. 

\begin{figure}
\plotone{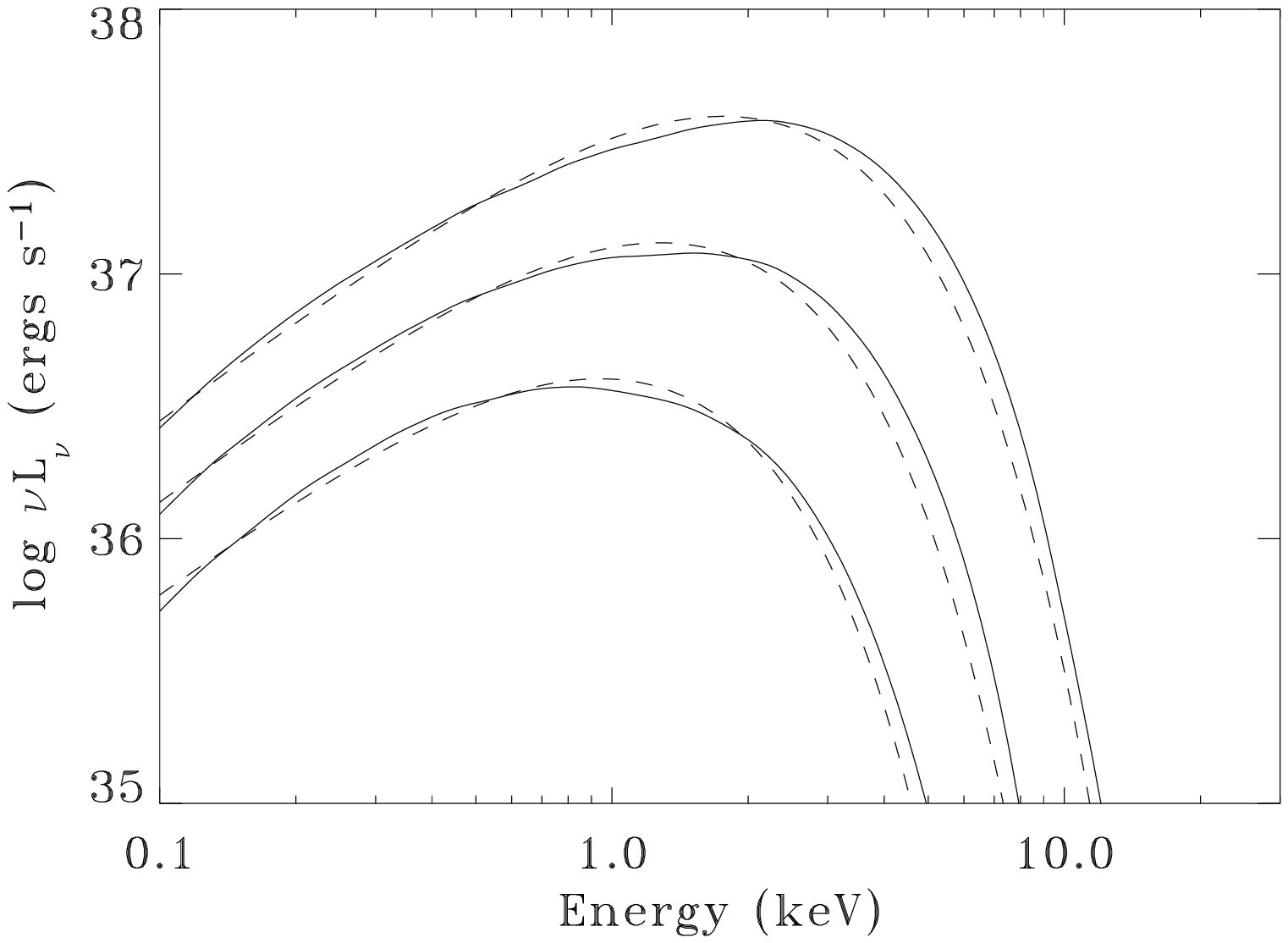}
\caption{
Integrated disk SED's for three values of $l$ (solid) observed from an 
$i=45^{\circ}$ by an observer at infinity.  At each $l$, we also plot the best-fit,
fully relativistic disk models in which the local flux is assumed to be an 
isotropic diluted blackbody (dashed). Each model has
$\alpha=0.1$ and $a=0.998$ and from lower left to upper right the curves 
correspond to $l=0.01$, 0.03, and 0.1 with best fit $f_{\rm col}=1.4$, 1.41, and 1.47
respectively. The diluted blackbody spectra are plotted at 114\% ($l=0.01$) and 112\%
($l$=0.03 and 0.1)
of their intrinsic luminosity to account for limb darkening in the non-LTE
atmosphere models.
\label{fig:lkerr0.71}}
\end{figure}

Relativistic broadening also accounts for the hardening of the SED.
As $a$ increases from zero to 0.998, 
$r_{\rm ms}$ decreases from 6.0 to 1.23. Moving closer to the event horizon 
amplifies the relativistic effects
on the photon geodesics, altering the SED observed from infinity.
At $i=70^{\circ}$, strongly blue-shifted emission from matter moving
toward the observer pushes the high energy tail to larger frequencies. 
In Figure \ref{fig:lkerr0.71}, we plot the same set of models as viewed from
a lower inclination of $i=45^{\circ}$.  The relativistic broadening is clearly
reduced and both the spectral peak and the high energy tail occur at lower
energies. 

In the hot inner annuli, the metals are highly ionized. Even for iron, a 
substantial fraction of the atoms are completely stripped of electrons
in the $l=0.1$ model. This reduces the bound-free opacity and electron
scattering plays a more dominant role than in the Schwarzschild case.
As a result, the diluted blackbody models which best approximate the solid 
curves generally have higher values of $f_{\rm col}$ and the breaks near the 
metal edges are not as prominent.  A calculation for these hotter annuli finds
$y^{\ast}_{\nu} \gtrsim 1$ for most relevant photon energies in the $l=0.03$ 
and 0.1 models.  The agreement between the diluted blackbody and atmosphere 
model SED's improves with increasing $l$. For $i=70^{\circ}$, the maximum 
deviations are always less than or equal to those in the corresponding 
Schwarzschild model. Since bound-free opacity is reduced, the ratio of 
absorption to scattering is much more frequency dependent for the higher
luminosity models and the improved agreement between the two types of spectra
is likely due to Comptonization.  The local limb darkening at characteristic
photon energies emitted by the hot inner annuli is even better approximated 
by that of a semi-infinite Thompson scattering atmosphere  (Chandrasekhar 1960)
than the Schwarzschild case depicted in Figure \ref{fig:limbd}.

\subsection{Dependence on $\alpha$ \label{alpha}}

As discussed in section \ref{method}, the set of three parameters 
$m_{0}$, $T_{\rm eff}$, and $Q$ along with the abundances and dissipation 
profile,
uniquely determine the structure and spectrum of an annulus in our models.
Of these quantities, only $m_0$ depends on $\alpha$. If all other
quantities are fixed, an increase (decrease) in $\alpha$ produces a larger 
(smaller) stress.  For the accretion rate to remain constant, there must be a 
corresponding reduction (increase) in the surface density and therefore in 
$m_0$. In a constant density, radiation pressure dominated one-zone model, a
change in $m_0$ leads to an equivalent fractional change in the density 
because the scale height is independent of $\alpha$ (SS73).

\begin{figure}
\plotone{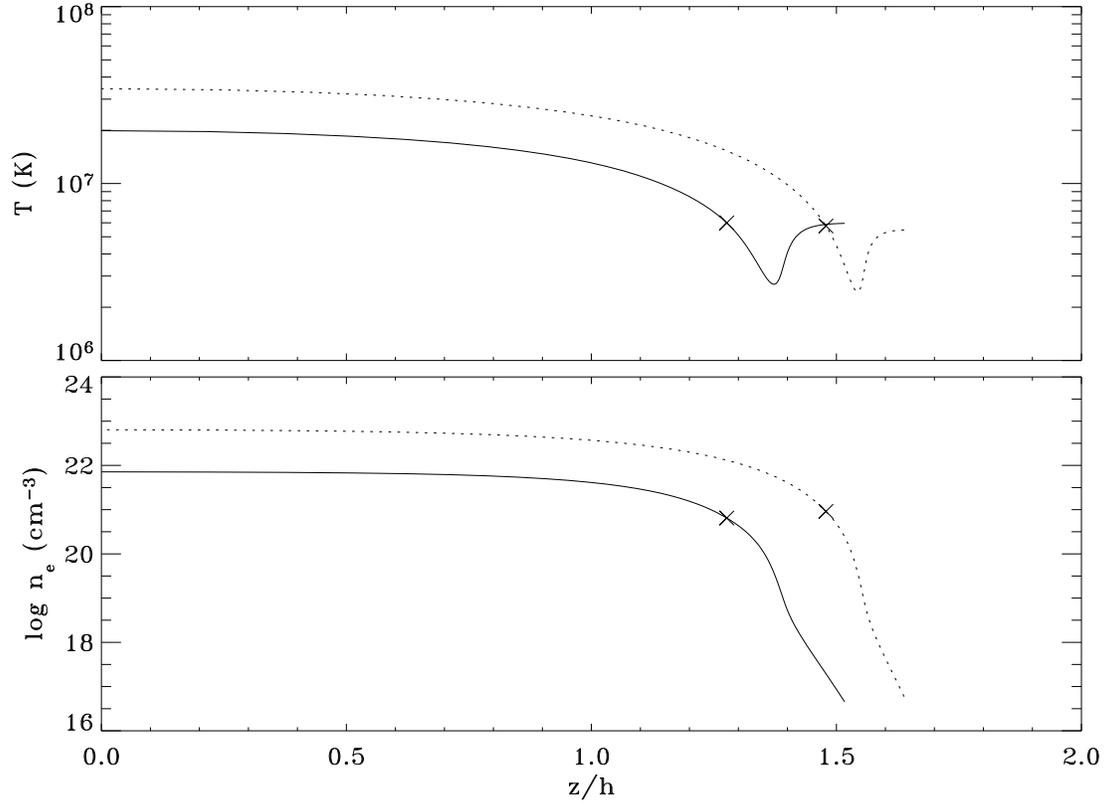}
\caption{
Temperature (top) and electron number density (bottom) as a function of 
height above the midplane. The atmospheres are located at $r=12.6$ in a disk
accreting on to a Schwarzschild black hole with $l=0.1$ and $\alpha=0.1$
(solid) or $0.01$ (dotted).  The x's mark the position of $\tau^{\ast}_{\nu}$
evaluated for photon frequencies near the peak
($\nu_{\rm peak} \approx 5 \times 10^{17} \, \rm Hz^{-1}$) of the SED.  The
scale height, $h$, is evaluated using a radiation pressure dominated one-zone
calculation (HH98 eq. 53).
\label{fig:alpdepth}}
\end{figure}

In Figure \ref{fig:alpdepth} we compare the vertical temperature and
electron number density $n_e$ profiles in disks with different values of
$\alpha$.  The solid and dashed curves correspond to 
$\alpha=0.1$ and 0.01 respectively. We plot results for annuli at $r=12.6$
in disks with $a=0$ and $l=0.1$. Both of these atmospheres are expected to
be radiation pressure dominated so we scale the $z$ coordinate with the 
appropriate one-zone model $h$ (HH98 eq. 53). The x's in Figure 
\ref{fig:alpdepth} mark the location of $\tau^{\ast}_{\nu}$ evaluated at a 
frequency near the spectral peak 
($\nu_{\rm peak} \approx 5 \times 10^{17} \, \rm Hz^{-1}$).

Below one scale height the
models are consistent with the predictions of a one-zone model.
The densities are nearly constant and $n_e$ in the $\alpha=0.01$ 
atmosphere is about a factor of ten larger than in the $\alpha=0.1$ case.
The flux rises linearly with $z$ providing a force that balances gravity.
However, above $z=h$ the flux rises less steeply
and begins to asymptote to $\sigma T_{\rm eff}^4$.  Gas pressure gradients
are therefore needed to
balance the continued increase in gravity and the density decreases
exponentially.  The x's indicate that most of the spectrum is formed where
the density is dropping rapidly and not where the density is relatively
constant.  Disk models which neglect vertical gradients in density
will therefore overestimate the ratio of absorption to scattering opacity
and underestimate the effective Compton $y$-parameter.

\begin{figure}
\plotone{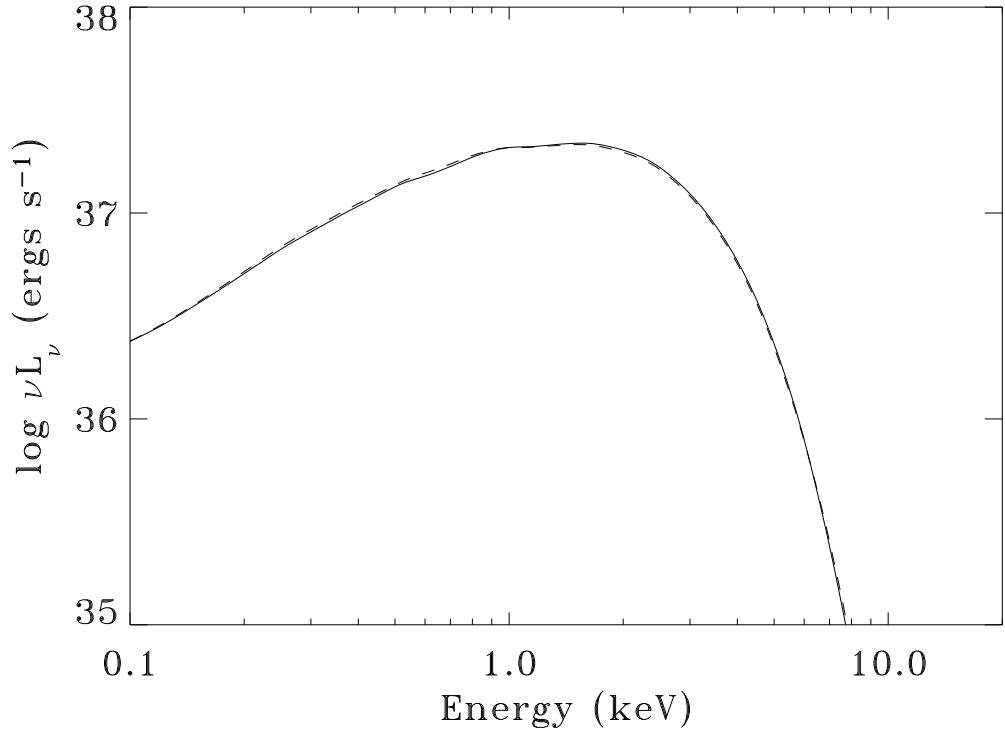}
\caption{
Integrated SED's observed at $i=70^{\circ}$ by 
an observer at infinity for disk with $l=0.1$ and $a=0$.  The curves with
$\alpha=0.1$ (solid) and 0.01 (dashed) are nearly indistinguishable.
\label{fig:alpspec}}
\end{figure}

Despite their differences for $z\lesssim h$, the temperature and density
profiles seen from the surface down to $\tau^{\ast}_{\nu}$ are nearly identical
in the two annuli.  In other annuli at larger $r$ and smaller $T_{\rm eff}$,
the $\tau^{\ast}_{\nu}$ surfaces move further from the midplane so that regions
of spectral formation remain near the surface where the density decreases 
exponentially.  In general, the temperature, density and ion level population
profiles of the two models still do not differ significantly in these surface
regions at any $r$. 
In Figure \ref{fig:alpspec} we compare the integrated SED's from these disks
viewed at $i=70^{\circ}$. They are nearly indistinguishable.

\section{The Stress Prescription}
\label{stress}

Most of the standard model disks ($\alpha$-disks) presented in 
section \ref{std} are
radiation pressure dominated in the innermost annuli.  Such disks have
long been known to be ``viscously'' and thermally unstable.  One way to avoid
these instabilities is to construct models (hereafter $\beta$-disks)
in which the accretion stress scales with gas pressure $P_{\rm gas}$ 
rather than total pressure $P$ in equation (\ref{eq:stress}) 
(Lightman \& Eardley 1974). It has been argued that such a scaling could be the
consequence of magnetic field buoyancy limiting magnetic stresses
(Sakimoto \& Coroniti 1981; Stella \& Rosner 1984).

As discussed in section \ref{alpha}, altering the stress prescription only
changes $m_0$ in a one-zone model.  A process that limits
the magnetic stress (e.g. magnetic field buoyancy) may also alter the vertical
dissipation profile, but we discuss the impact of the dissipation profile 
separately in section \ref{dissip}.  For fixed $\alpha$, replacing $P$
by $P_{\rm gas}$ in equation (\ref{stress}) results in a lower stress.
Thus, a $\beta$-disk requires a larger $m_0$ to produce the same accretion rate
as an $\alpha$-disk.  The larger $m_0$ results in greater midplane densities
and temperatures.

\begin{figure}
\plotone{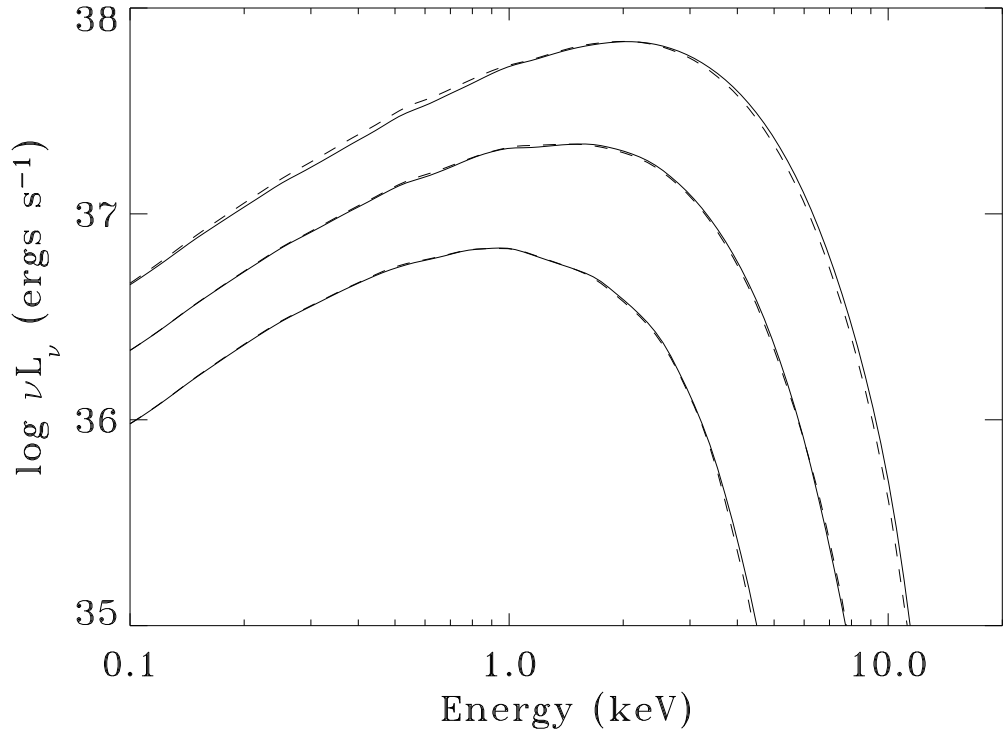}
\caption{
Integrated disk SED's for three values of $l$ observed at
$i=70^{\circ}$ by an observer at infinity. Each model has
$\alpha=0.1$ and $a=0$ and from lower left to upper right the curves 
correspond to $l=0.03$, 0.1, and 0.3.  The $\alpha$-disks (solid) are 
equivalent to the curves plotted in Figure \ref{fig:lspec}.  The
$\beta$-disks (dashed) show little deviation from the $\alpha$-disks.
\label{fig:beta0.34}}
\end{figure}

We construct $\beta$-disk models with $\alpha=0.1$ for Schwarzschild black 
holes with $l=0.03$, 0.1, and 0.3 and maximally spinning Kerr
black holes with $l=0.03$ and 0.1. At $l=0.01$ the $\alpha$-disk models are 
gas pressure dominated down to $r_{\rm in}$ and are therefore equivalent to 
$\beta$-disks.
We calculate $m_0$ by integrating the LTE-grey model until equation 
(\ref{eq:stress}) is satisfied for $P=P_{\rm gas}$.
In Figure \ref{fig:beta0.34}, we compare the integrated SED's of the
Schwarzschild models with the equivalent $\alpha$-disk spectra for an 
observer at $i=70^{\circ}$.  The results for the maximally spinning Kerr
models are similar.  We find that the stress prescription has little 
effect on the black hole disk SED's for the same reasons as in section 
\ref{alpha}. 
Specifically, $\tau^{\ast}_{\nu}$ lies near the surface of the disk where 
the vertical structure of the two types of annuli are always similar regardless
of the conditions at the midplane.  Photons emitted at $\tau^{\ast}_{\nu}$
in the $\beta$-disks encounter nearly identical profiles of density, 
temperature, and level populations as those in an $\alpha$-disk.

\section{Torques on the Inner Boundary}
\label{torque}

The no-torque boundary condition may be invalid if magnetic fields
keep the material in the plunging region causally connected to the disk
(Gammie 1999; Krolik 1999). In that case, the surface density and brightness 
of the disk are modified from the standard model.  The corresponding changes 
in $T_{\rm eff}(r)$ and $m_0(r)$
can be parametrized in terms of $\Delta \eta$, the change in efficiency
of converting energy into radiative flux (Agol \& Krolik 1998).
For a Schwarzschild black hole with no inner torque, $\eta=0.057$.
We calculate a torqued Schwarzschild disk with $l=0.1$, $\alpha=0.1$,
and $\Delta \eta=0.05$ so that the efficiency is nearly double the 
no-torque case.  Since we fix the luminosity of the two disks, the accretion
rate of the torqued disk is lower. Our method of
integrating the LTE-grey model fails to converge for some of the hottest 
annuli so we use the one-zone calculation to find $m_0$ throughout the disk.
We computed the values of $\alpha$ required to satisfy equation
\ref{eq:stress} in the converged non-LTE annuli, and they differed from
the assumed value by less than 50 percent.
For the reasons discussed in section \ref{alpha}, we do not expect the
the resulting spectrum to be sensitive to this small discrepancy.
In the disk without a torque, $T_{\rm eff}$ increases with decreasing
$r$ until reaching a maximum at $r_{\rm max}> r_{\rm ms}$ and then falls to
zero at $r_{\rm ms}$.  For the torqued disk, $r_{\rm max}=r_{\rm ms}$ and
$T_{\rm eff}$ continues to rise all the way down to the inner edge of the disk.
Despite the decrease in $\dot m$, the fluxes in the innermost annuli are higher
than in a standard disk at the same luminosity.

\begin{figure}
\plotone{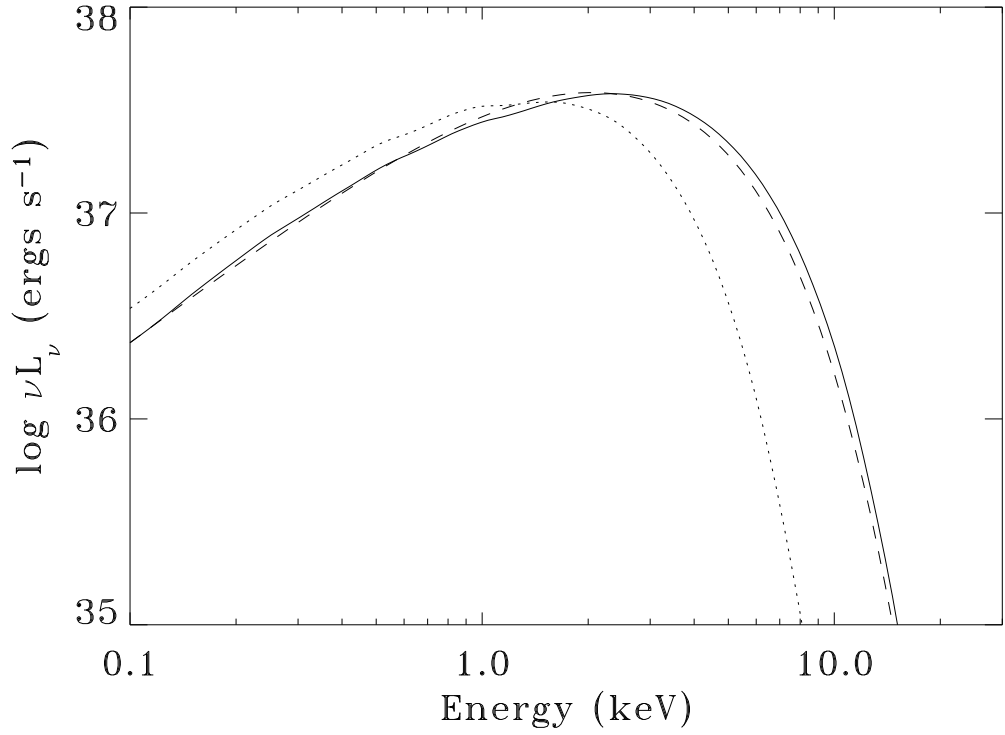}
\caption{
SED (solid) from a torqued disk with $l=0.1$, $a=0$, 
$\alpha=0.1$, and $\Delta \eta=0.05$ viewed at
$i=70^{\circ}$ by an observer at infinity. We also plot the best-fit, fully 
relativistic disk models in which the local flux is assumed to be an 
isotropic diluted blackbody with $f_{\rm col}=1.61$ (dashed).
The diluted blackbody spectra are folded through the same transfer function and viewed 
at the same inclination as the solid curve.  The SED of the torqued disk
is considerably harder than the SED of of disk without a torque (dotted).
\label{fig:torque}}
\end{figure}

In Figure \ref{fig:torque}, we compare the
SED's of the torqued (solid) and standard model (dotted) disks viewed at $i=70^{\circ}$.
Using the same surface brightness and folding through the same
transfer function, we also calculate spectra assuming the
local emission produces an isotropic diluted blackbody with a best-fit 
$f_{\rm col}=1.61$.
Owing to the larger fluxes, the innermost annuli of 
the torqued disk are hotter and the SED is much harder.  In these
annuli, the metals are highly ionized and provide little opacity.  
Comptonization is saturated leading to annuli with Wien-like spectral profiles.
The diluted blackbody is still good to about 10\% for frequencies at or below
the peak. The torqued 
disk requires a slightly larger $f_{\rm col}$ than the corresponding standard
model disk.

One motivation for examining torqued disks is the hope that they might be
well approximated by a {\it diskbb} model.
Since the {\it diskbb} model is relatively inexpensive to calculate it is 
particularly useful for fitting data.   
It assumes a simple surface brightness profile which
ignores the no-torque inner boundary condition and the relativistic
corrections to the disk structure equations.  The surface brightness of the
torqued disk provides a better match to this profile so we fit
our fully relativistic models with the {\it diskbb} model.  Fits for
the disk models with no inner torque are discussed in section \ref{lvst}.

We perform the fits by first generating XSPEC table models from our 
model SED. Next, we use these table models along with the {\it XMM-Newton} 
European Photon Imaging Camera pn (EPIC-pn) response matrices to create 
artificial PHA data sets.  We also account for the intervening
Galactic absorption with a neutral absorber model.  Accounting for 
photoelectric absorption
is important because it affects the statistical weight of the bins used in
the fits.  The absorption reduces the relative fraction of photon counts on 
the low energy end of the EPIC-pn band and makes the fit more sensitive to the
emission at higher energies.  We choose a H column of 
$5 \times 10^{21} \rm cm^{-2}$ as a representative value for a typical
X-ray binary source. We also assume the source is at 5 kpc with 75 ksec 
exposure, providing a high signal-to-noise ratio dataset.  We then fit these 
data with a {\it diskbb} model over the 
0.3-10 keV energy range including the same absorption used to generate the
artificial spectrum.  The reduced $\chi^2$ of the fit is poor with 
$\sim 20\%$ deviations
below 1 keV and $\sim 10\%$ differences near the spectral peak.  The best-fit
{\it diskbb} model is narrower than our model, providing a deficit of flux
at the high and low energy ends of the band and an excess near the spectral
peak. This is in part due to the departures from blackbody emission, but is
due mostly to relativistic transfer effects and the remaining 
differences in the surface brightness profile.  Considering a larger torque
would increase the similarity in the surface brightness profiles. This should
improve the quality of the fit but discrepancies would remain.  Even when
we assume locally isotropic blackbody emission with a surface 
brightness identical to the {\it diskbb} model, the resulting SED is
intrinsically broader than the {\it diskbb} due to the effects of
relativistic transfer.

\section{The Dissipation Profile}
\label{dissip}

In section \ref{method} we stated that our standard model assumes that the
energy dissipation rate per unit volume $\epsilon$ is locally proportional 
to the density $\rho$. Before exploring other dissipation profiles, it is 
worthwhile reminding the reader
what this implies for a radiation pressure dominated disk.
In that case, hydrostatic balance reduces to
\be
\frac{-1}{\rho}\frac{d P_{\rm rad}}{d z}=\frac{\kappa_{F} F}{c}=
\frac{G M}{R^3}\frac{C}{B} z
\ee
where $P_{\rm rad} \sim P$ is the radiation pressure, $C$ and $B$
are general relativistic correction factors (Riffert \& Herold 1985), and
$\kappa_{F}$ is flux mean opacity.  In these scattering dominated disks,
$\kappa_{F} \approx \kappa^{\rm es}$ which is only weakly dependent on
$z$ through the free electron fraction.
Therefore  $F \propto z$, implying that $\epsilon$ is 
constant when the gradient in radiation pressure balances gravity.
Our assumption that $\epsilon\propto\rho$ then implies that the
density is constant so long as gas pressure gradients are negligible.
From the definition of column mass $dm=-\rho dz$, it is easy to see that our
assumption reduces to
\be
\label{eq:dis}
\epsilon=\rho \left(-\frac{d F}{d m}\right)=\rho \frac{\sigma T^4_{\rm eff}}{m_0}
\ee
where $\sigma$ is the Stefan-Boltzman constant.  The first equality holds by
definition and the second follows from our assumption which implies
$d F/d m$ is independent of $m$.  We have set
$F(m)=\sigma T^4_{\rm eff}(1-m/m_0)$ in all the models discussed above.

\subsection{Physically Motivated Dissipation Profiles}
\label{sims}

We would like to examine how the choice of $F(m)$ affects the
vertical structure of the accretion disk but there is no firm theoretical
basis to guide our analysis.  It is commonly accepted that the 
magnetorotational instability (MRI, Balbus \& Hawley 1991) can provide a means
for tapping the free energy of the differential rotation and producing 
turbulence in which energy may cascade irreversibly from large to small 
scales.  However, the nature of non-radiative vertical energy transport 
and the mechanisms for dissipation in these disks are still not well 
understood. Despite these uncertainties, there is a rather
robust prediction of magnetic field buoyancy in radiation 
dominated annuli (Sakimoto \& Coroniti 1981; Stella \& Rosner 1984).
It is therefore quite reasonable to expect that a larger fraction of the
dissipation occurs nearer the surface than is assumed in equation \ref{eq:dis}.

Numerical simulations which include radiative diffusion 
in a vertically stratified disk (Turner 2004) bear out this
prediction. The radiation-MHD equations are solved for a
patch of disk centered at 200 $R_{\rm g}$ from a $10^8 \, M_{\odot}$ black
hole. Magnetic field energy is produced fastest by the MRI two to three 
density scale heights away from the midplane.
The magnetic field is buoyant and rises toward the surface at
approximately the Alfv\'en speed.  The 
density is more centrally concentrated than in a standard model disk with
the same surface density and radiative flux.  The time averaged dissipation
is not at all consistent with the $\epsilon \propto \rho$ assumption as much
of the dissipation takes place in the low density surface layers.

Similar calculations are not yet available for disks accreting onto 
$10 M_{\odot}$ black holes. In order to explore the possible consequences of
magnetic buoyancy,
we choose a form for $d F/d m$ which approximates the Turner (2004) results 
for a  $10^8 M_{\odot}$ black hole.  We then use this form to calculate $F(m)$
with choices of $m_0$ and $T_{\rm eff}$ appropriate for an annulus around
a $10 M_{\odot}$. Although these simulations are the best quantitative
model for $F(m)$ available, this choice should not 
be considered overly restrictive. There is still considerable uncertainty
in the simulation itself. Although 29\% of the dissipation is due to Silk 
damping (Agol \& Krolik 1998), most is still the result of 
numerical reconnection. Furthermore, the outermost 
grid zone is quite optically thick but our models extend to $\tau \ll 1$ so 
there are several decades in $m$ for which $F(m)$ is not constrained by the 
simulation results. Following HH98, we approximate
$d F/d m$ as a broken power law with $m$ being the dependent variable.  
The broken power law is 
parametrized by a division point $m_{\rm d}=f_{\rm d} m_0$ and
exponents above ($\zeta_1$) and below ($\zeta_0$) the division.  For the
standard model, $\zeta_0=\zeta_1=0$.
In the simulation, the dissipation and density
are functions of position and time so we average the domain horizontally
and we average over time from  20 to 40 orbits.  
The broken power law
provides a reasonable approximation with $\zeta_0=-0.9$, $\zeta_1=0$, 
and $f_{\rm d}=0.004$ so we do not consider more complicated profiles.   

Next we insert this choice of $d F/d m$ in an annulus with $m_0$ and 
$T_{\rm eff}$ appropriate for $r=12.6$ in a $l=0.1$, $a=0$, $\alpha=0.1$, and 
$10 M_{\odot}$ standard model disk.  We had difficulties obtaining convergence
for non-LTE models in which the Eddington factors were updated after each
solution of the radiative transfer equation. If we instead fix the Eddington 
factors after the first radiative transfer solution, the models converge.  We 
believe the lack of convergence in the first case is a result of our 
calculation method rather than an instability in the model. Furthermore, we 
doubt the assumption in the second case has 
a strong effect on the resulting spectra. The maximum fractional change in 
the Eddington factors after a global solution of the radiative transfer is 
less than 3\% for all frequencies and depths and is generally significantly
smaller.

The resulting annulus structure (dotted curve) is compared with the 
corresponding standard model (solid curve) in Figure \ref{fig:dissip}.  The
two are quite different.  The
radiative flux is plotted in the top panel.  For $z < h$, the solid curve
rises linearly because the radiative flux must balance the linearly increasing
vertical gravity.  The dotted curve falls below this line so the gradient
in the gas pressure also provides significant support against gravity in the
modified annulus. The temperature is shown in the middle panel.  The surface
layers above $\tau=1$ (marked by the squares) are somewhat hotter in the
modified annulus owing to the increased fraction of dissipation in that
model.  The
electron number density is shown in the bottom panel.  The density is
more centrally concentrated than in the standard model annulus as a density
gradient is necessary to compensate for the reduced radiation pressure
support.  The temperatures
and densities at $\tau^{\ast}_{\nu}$ (marked by x's) are nearly equal
in the two models, though the shape of the temperature and density profiles 
are more dissimilar from each other
than in the models with differing stress prescriptions.  

The specific intensities at $i=55^{\circ}$ are plotted in Figure 
\ref{fig:disspec}.  The SED of the modified annulus is slightly harder giving
an $f_{\rm col}$ which is about 10\% higher and the He-like Fe
absorption edge at 8.8 keV becomes an emission edge.  Despite these 
differences, the model spectra are still qualitatively quite similar.  The 
strong temperature inversion in the modified annulus is above the photosphere
for frequencies redward of the 8.8 keV edge and has little effect on the resulting
SED.

\subsection{Dissipation and Coronae}
\label{corona}

\begin{figure}
\plotone{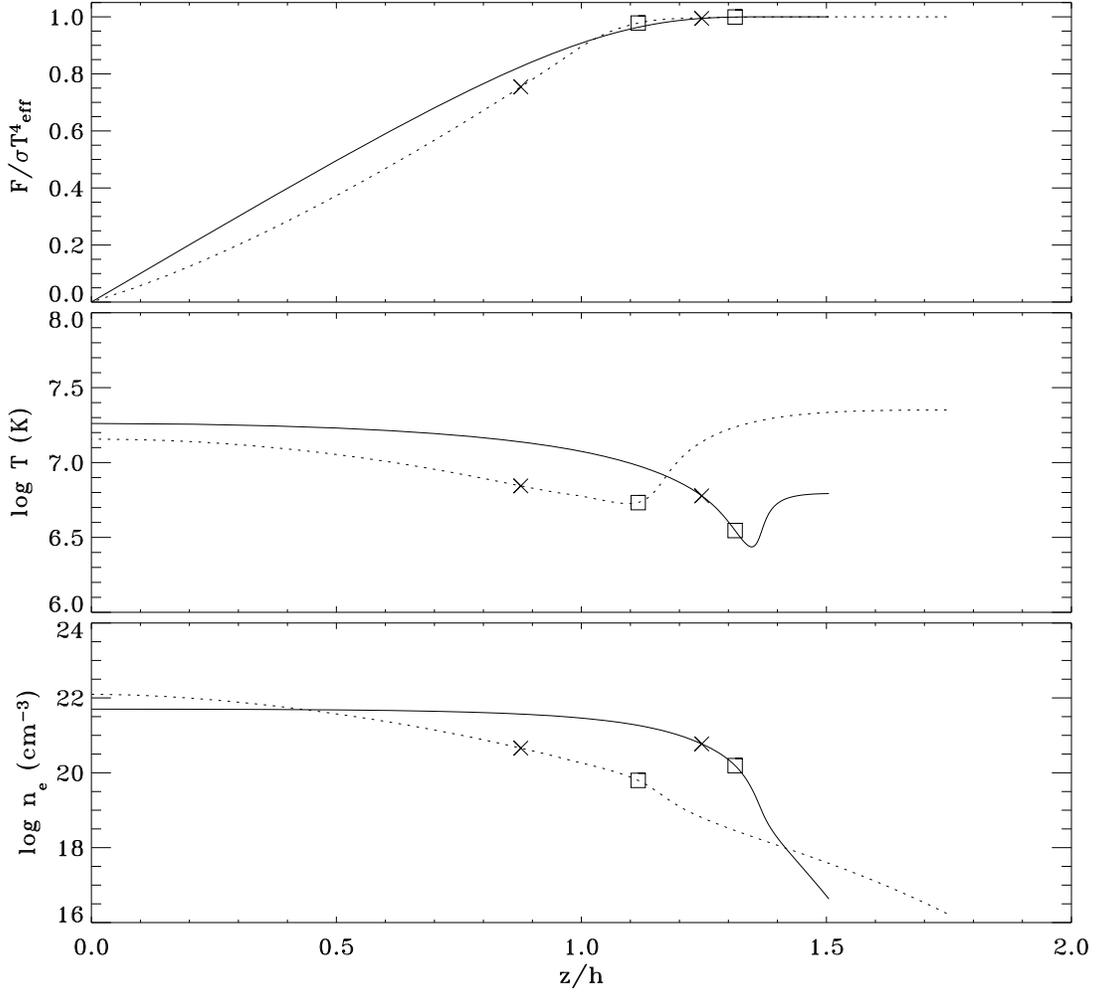}
\caption{
Radiative flux (top), temperature (middle) and electron number density 
(bottom) as a function of 
height above the midplane. The atmospheres have $Q$, $m_0$, and $T_{\rm eff}$
appropriate for $r=12.6$ in a disk
accreting on to a Schwarzschild black hole with $\alpha=0.1$.  We compare
the standard model ($\zeta_0$=$\zeta_1$=0; solid curve) to a model with a
modified dissipation profile ($\zeta_0$=-0.9, $\zeta_1$=0,$f_{\rm d}=0.004$;
dotted curve). 
The squares mark the position of the Thompson photosphere and the x's mark 
$\tau^{\ast}_{\nu}$ evaluated for photon frequencies near the peak
($\nu_{\rm peak} \approx 5 \times 10^{17} \, \rm Hz^{-1}$) of the SED.  The
scale height, h, is evaluated using a radiation pressure dominated one-zone
calculation (HH98 eq. 53).
\label{fig:dissip}}
\end{figure}

\begin{figure}
\plotone{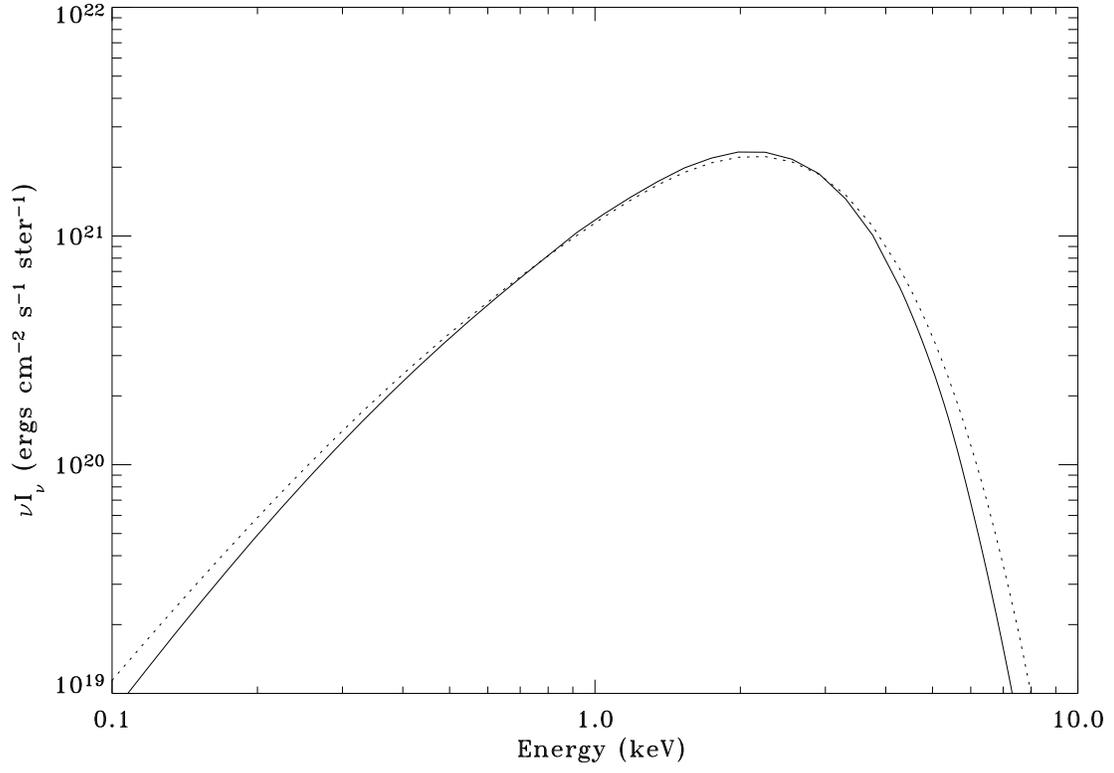}
\caption{
Specific intensity viewed at
$i=55^{\circ}$ in the local frame for the annuli shown in Figure
\ref{fig:dissip}. The models parameters are $\zeta_0$=$\zeta_1$=0 
(solid curve), $\zeta_0$=-0.9, $\zeta_1$=0, and 
$f_{\rm d}=0.004$ (dotted curve).
\label{fig:disspec}}
\end{figure}

The geometry and origin of the coronae in accretion disks remain uncertain.
One plausible explanation is that the coronae are due to hot electrons
heated by reconnection
of magnetic fields in the optically thin layers at the disk surface.  This
is the essence of the Svensson \& Zdziarski (1994) model where it is assumed 
that some fraction of released energy is transported without dissipation to a 
corona above the disk. It can be seen from the 
top panel of Figure \ref{fig:dissip} that very little 
dissipation occurs above the Thompson photosphere in the models described 
in \ref{sims}. We have considered several different prescriptions for $F(m)$ 
in which larger fractions of the dissipation occur at low $m$. The choice of
a broken power law for $dF/dm$ provides a smooth, continuous dissipation
profile which is different from the Svensson \& Zdziarski (1994) model
in which $dF/dm$ would be a step function. 

\begin{figure}
\plotone{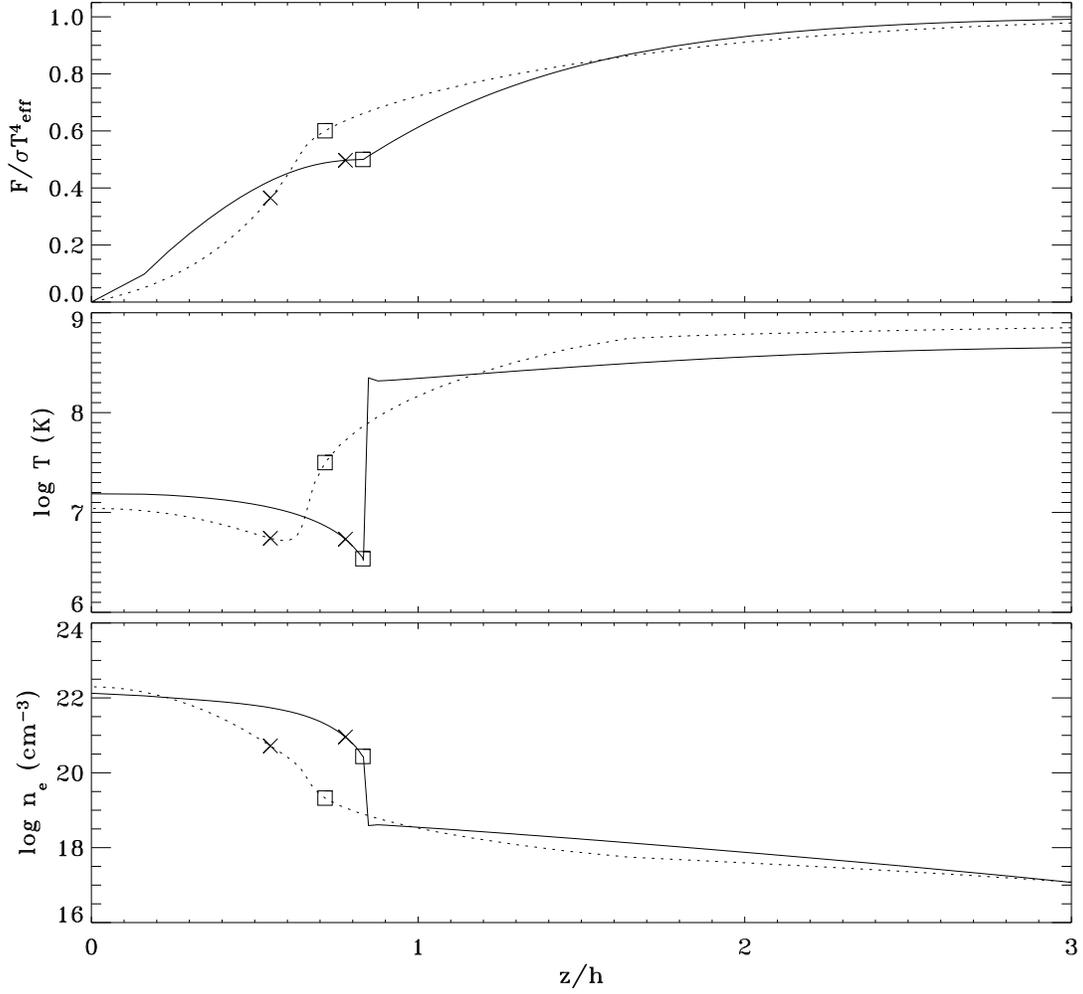}
\caption{
Radiative flux (top), temperature (middle) and electron number density 
(bottom) as a function of 
height above the midplane. The atmospheres have $Q$, $m_0$, and $T_{\rm eff}$
appropriate for $r=12.6$ in a disk
accreting on to a Schwarzschild black hole with $\alpha=0.1$.
The first model (solid curve) is similar to a standard model annulus
($\zeta_0=\zeta_1=0$) except that $dF/dm$ is not continuous and 50\% of the flux 
is dissipated above the Thompson photosphere. The second model (dashed curve)
has a continuous $dF/dm$ with $\zeta_0$=-1.1, $\zeta_1$=0, and 
$f_{\rm d}=0.0001$. About 40\% of the dissipation occurs above Thompson depth
unity.
The squares mark the position of the Thompson photosphere and the x's mark 
$\tau^{\ast}_{\nu}$ evaluated for photon frequencies near the peak
($\nu_{\rm peak} \approx 5 \times 10^{17} \, \rm Hz^{-1}$) of the SED.  The
scale height, h, is evaluated using a radiation pressure dominated one-zone
calculation (HH98 eq. 53).
\label{fig:corona}}
\end{figure}

In Figure \ref{fig:corona}, we show two annuli in which almost half
of the dissipation occurs above the Thompson photosphere. The first case
(solid curve) is similar to a Svensson \& Zdziarski (1994) model in that 
$dF/dm$ is constant below the photosphere as in the standard model disks, but
$dF/dm$ is discontinuous at the photosphere. Nearly 50\% of the dissipation
occurs above a Thompson depth of unity and the resulting annulus has a sharp rise
(drop) in the temperature (density) just above the scattering photosphere. In order
to better approximate these gradients, we increase our number of depth points but
memory restrictions then limit us to just considering H, He, C, N, O, and Fe for
this annulus only. The
neglect of less abundant species seems to have little effect on the vertical
structure or spectrum. In the second case (dotted curve)
$dF/dm$ is a continuous broken power law with $\zeta_0$=-1.1, $\zeta_1$=0, and 
$f_{\rm d}=0.0001$. About 40\% of the dissipation occurs above the Thompson
photosphere and another $\approx 20\%$ occurs above $\tau^{\ast}_{\nu}$
(measured near the peak). We consider the same
$m_0$ and $T_{\rm eff}$ used for the models plotted in the Figures
\ref{fig:dissip} and \ref{fig:disspec}. The profiles in the discontinuous model
closely resemble those of a standard model disk below the photosphere, but above
there is a sharp rise in the temperature and drop in the density.  In the 
second case,
the shape of the density and temperature profiles are similar to the 
modified annulus discussed in section \ref{sims} but on a more exaggerated
scale.  The surface layers are much more extended, the surface temperature
and central density are higher, and the densities at the surface are lower.

\begin{figure}
\plotone{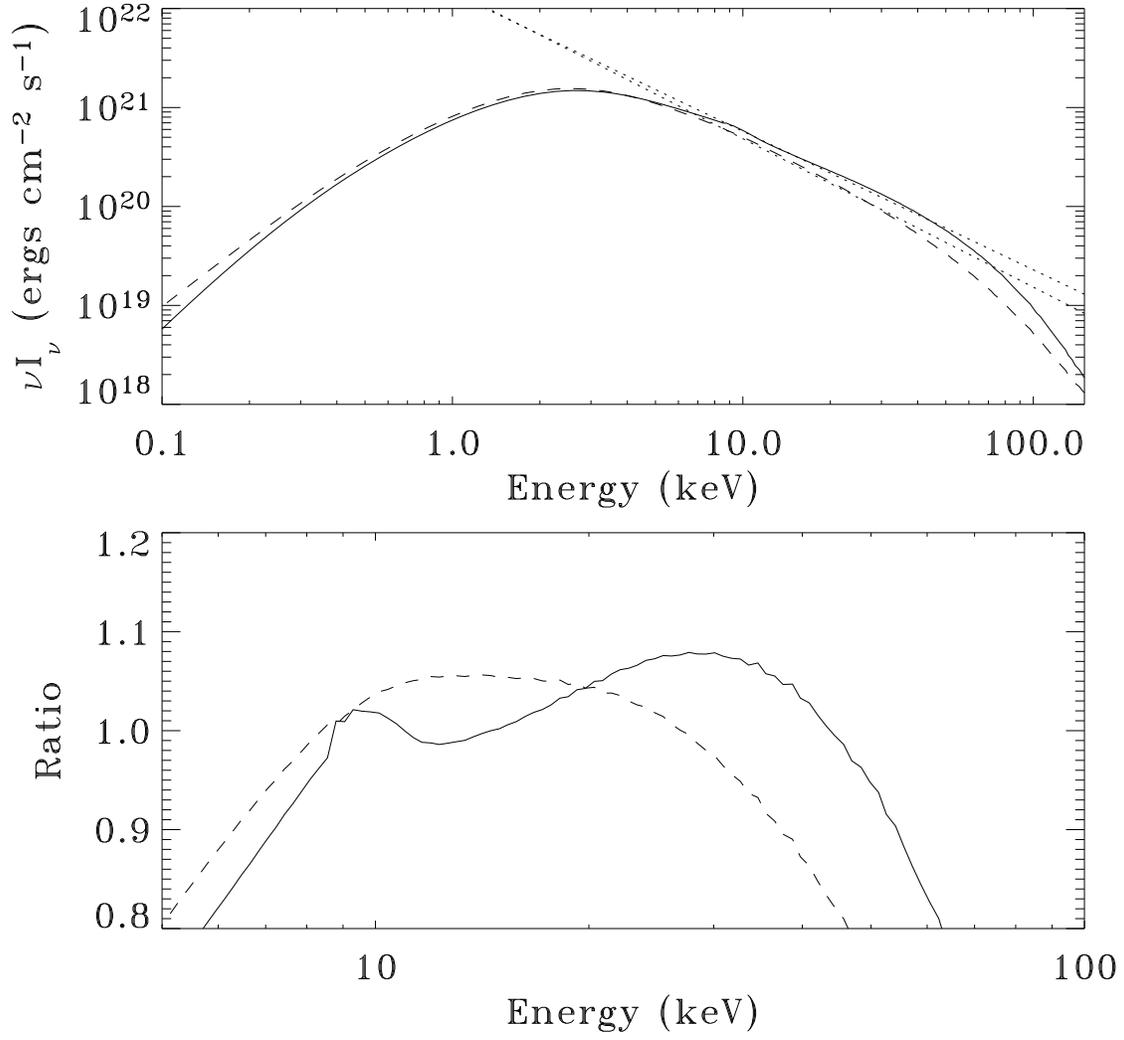}
\caption{
Specific intensity viewed at
$i=55^{\circ}$ in the local frame for the annuli shown in Figure 
\ref{fig:corona}. 
The first model (solid curve) is similar to a standard model annulus
($\zeta_0=\zeta_1=0$) except that $dF/dm$ is not continuous and 50\% of the flux 
is dissipated above the Thompson photosphere. The second model (dashed curve)
has a continuous $dF/dm$ with $\zeta_0$=-1.1, $\zeta_1$=0, and 
$f_{\rm d}=0.0001$
\label{fig:corspec}}
\end{figure}

We plot the SED's of these annuli in Figure \ref{fig:corspec}.  In the top
panel we plot the specific intensity at $55^{\circ}$.  In both cases,
Compton scattering by the hot thermal electrons above the photosphere produces
a steep tail with emission up to $\sim 100$ keV.  These tails are only crudely
approximated by a power law as there is curvature at low energies near the
peak and at high energies due to the thermal cutoffs.  The discontinuous model
(solid curve) produces a slightly harder spectrum.  A flatter, more
power law-like tail such as is commonly observed could presumably be produced by
choosing a more top heavy
$F(m)$, but the surface temperatures become too large and our treatment of
Compton scattering is no longer valid. We also include power law curves with 
photon 
indices of $\Gamma=3.4$ and 3.5 to approximate the two spectra over the
$\sim5-50$ keV band. 
In the bottom panel we plot the ratio of the model SED's to these power laws.
For the discontinuous model there is curvature due to the reduction in the
scattering albedo from bound-free opacity 
above $\sim 10$ keV and the spectrum resembles the putative Compton
reflection hump.  The high energy side of the hump is due to
both Compton down scattering by the cooler electrons near the photosphere
and the thermal cutoff of the hot electrons. There is a strong emission feature
at the He-like Fe edge.  This feature is likely enhanced because our treatment
only allows recombinations to the ground state.  Some fraction of this power would
be shifted to lower energies if we had considered recombinations to excited levels.
In the second case, the spectrum 
simply falls off at high energies and there is no structure in the residuals 
which could be characterized as a reflection hump.
A large fraction of dissipation occurs in the
spectral forming regions just below the photosphere so the temperature there
is high and the density is relatively low.  Almost all iron is completely 
ionized so that the bound-free opacity which produces the low energy side of the 
hump in the previous model is insignificant.

\section{Luminosity - Temperature Relation}
\label{lvst}

Detailed fits to individual observations are beyond the
scope of this paper.  Because we neglect irradiation of the disk surface, we
focus on comparison with the results of GD04.  Their sample is particularly 
relevant because they select observations in which the coronal emission is 
less than 15\% of the bolometric luminosity of the source.  

GD04 construct
plots of the temperature-luminosity relation for each source in their sample.
The luminosity of the disk model $L_{\rm disk}$ is plotted as a fraction of 
$L_{\rm Edd}$ versus the maximum color temperature $T_{\rm max}$. A major 
difficulty with inferring the physical properties of accretion disks is that 
the distance, inclination and black hole mass of the source are often not
all known with great accuracy.  Therefore, the vertical and horizontal position
of the locus of points on these curves is uncertain.  However, the shape of
the locus of points for an individual source should be robust to such 
uncertainties.

If the disk spectrum is well approximated by a multitemperature 
diluted blackbody
with fixed $f_{\rm col}$, $L_{\rm disk}$ should be proportional to  
$T_{\rm max}^4$.  Specifically, Gierli\'nski et al. (1999) assume that the
gravitational field may be approximated with a pseudo-Newtonian potential 
(Paczy\'nski \& Witta 1980) and find the relation
\be
\frac{L_{\rm disk}}{L_{\rm Edd}} \approx 0.583 \left(\frac{1.8}{f_{\rm col}}\right)^4 \left(\frac{M}{10 M_{\odot}}\right)\left(\frac{k T_{\rm max}}{1 \rm keV}\right)^4
\label{eq:lvst}
\ee
where $T_{\rm max}=T_{\rm eff,max} f_{\rm col}$.
GD04 find that most of the sources roughly follow this 
scaling, though several show some degree of relative hardening (increasing 
$f_{\rm col}$) as $L_{\rm disk}$ increases.

One obstacle to comparing with their results is that neither $T_{\rm max}$ nor
$f_{\rm col}$ is a well defined theoretical quantity in our models. 
In principle, we could use the $f_{\rm col}$ values from the best-fit
models shown in Figure \ref{fig:lspec}, but that fitting procedure is quite 
different from the method implemented by GD04.
In constructing their temperature-luminosity diagram, they fit the 
{\it diskbb} model to their spectra to account for the presumed disk
component. They then use the
best fit inner temperature and total model flux to calculate 
$L_{\rm disk}/L_{\rm Edd}$ and $T_{\rm max}$. In doing so they apply 
relativistic correction factors (Zhang, Cui \& Chen 1997) for a Schwarzschild
black hole at the appropriate inclination.  They also apply an additional
correction factor to account for the mismatch between the surface brightness
profile of the {\it diskbb} model and the pseudo-Newtonian potential
used to derive equation \ref{eq:lvst}.

To facilitate comparison, we try to reproduce this procedure as closely as
possible. We produce an artificial EPIC-pn PHA data set as described in section
\ref{torque}.  We also generate a second set of artificial spectra by the
same method but employing the {\it RXTE} Proportional Counter Array (PCA) 
response matrices.  Galactic absorption has less impact on the PCA photon
statistics so we do not include it in those artificial spectra and fits. 
The EPIC-pn spectra are useful because of
the high signal-to-noise and the low energy coverage while the
PCA spectra are necessary for comparing with GD04. Next,
we fit {\it diskbb} models over the 
0.3-10 keV and 3-20 keV energy ranges for the EPIC-pn and PCA data sets 
respectively. We also generate artificial spectra including a
{\it compTT} (Titarchuk 1994) component which is 10\% of the bolometric flux 
to account 
for possible Comptonized coronal emission.  These spectra are then fit with
a combined {\it diskbb} + {\it compTT} model. We find that the
best fit parameters of the {\it diskbb} models are only weakly sensitive to 
the presence or absence of the {\it compTT} component as long as we fit
for this component with another {\it compTT} model and {\it not} a power law.
Finally, we apply the same prescription outlined in GD04 to calculate
$L_{\rm disk}/L_{\rm Edd}$ and $T_{\rm max}$ from the model temperature and
flux.

\begin{figure}
\plotone{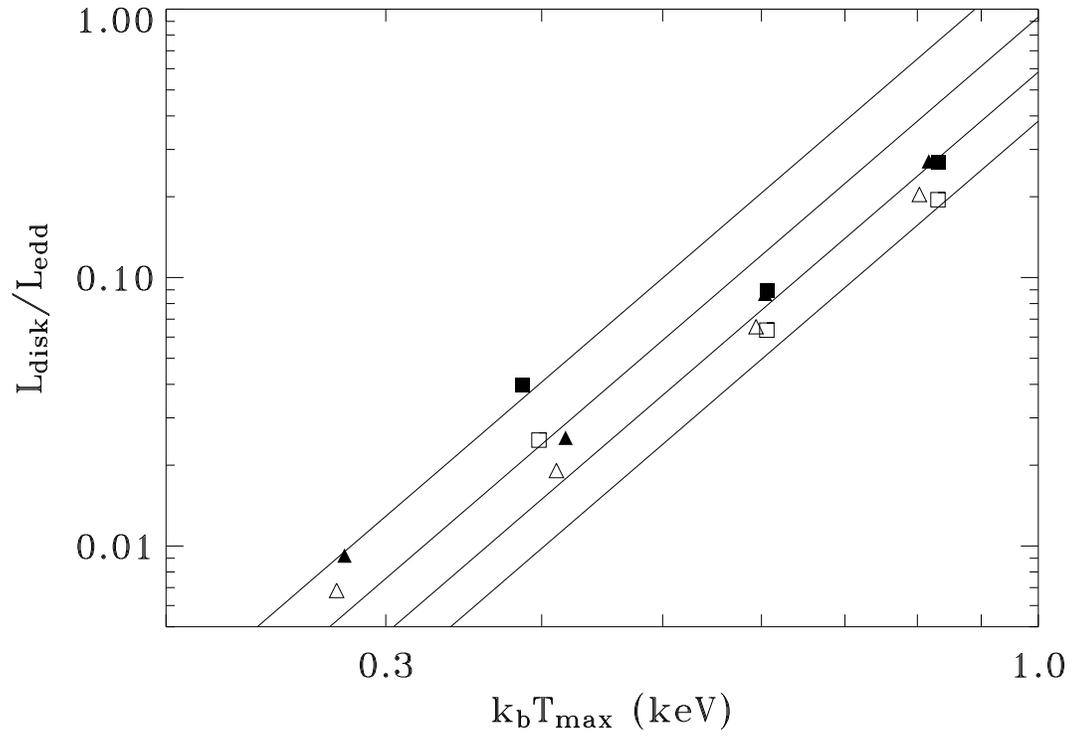}
\caption{
Luminosity-temperature relation measured for four Schwarzschild disks viewed 
at $i=45^{\circ}$ (filled symbols) and $70^{\circ}$ (open symbols).
The solid curves represent the lines of constant $f_{\rm col}=1.4$, 1.6,
1.8, and 2.0 (from top to bottom) from equation \ref{eq:lvst}. The triangles 
and squares
mark the {\it XMM} EPIC-PN  and {\it RXTE} PCA measurements respectively.
\label{fig:lvst}}
\end{figure}

\begin{figure}
\plotone{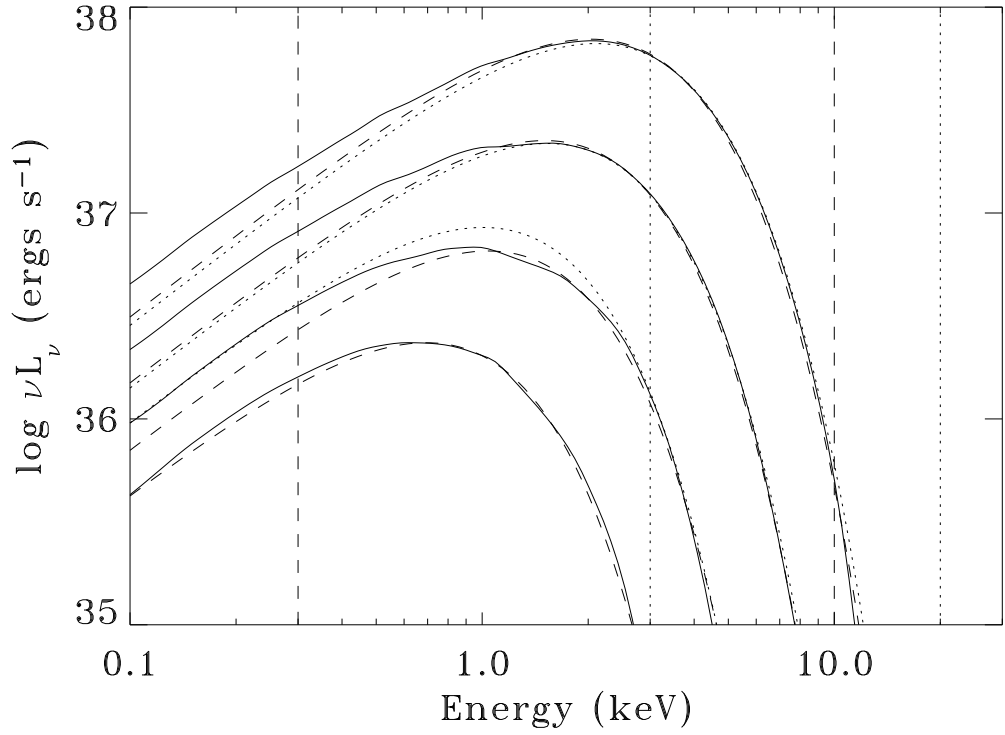}
\caption{
Integrated non-LTE disk SED's from Figure \ref{fig:lspec} (solid curve).  
At each $l$, we also plot the 
best-fit {\it diskbb} models for the {\it XMM-Newton} EPIC-pn (dashed) and
{\it RXTE} PCA (dotted) artificial data sets.  The vertical solid and dashed
lines mark the endpoints of EPIC-pn and PCA energy ranges respectively.
\label{fig:fits}}
\end{figure}

We show the results of this procedure for the four Schwarzschild models with
$l=0.01$, 0.03, 0.1, and 0.3 viewed at $i=45^{\circ}$ and $70^{\circ}$
in Figure \ref{fig:lvst}.  The solid curves represent the 
luminosity-temperature relation of equation \ref{eq:lvst} for
$f_{\rm col}=1.4$, 1.6, 1.8 and 2.0. The triangles and squares
mark the {\it XMM-Newton} EPIC-pn and the {\it RXTE} PCA measurements 
respectively. Only the highest three luminosities are plotted
for the PCA measurements because the model with $l=0.01$ emits only a small 
fraction of its luminosity above 3 keV and is not well constrained by a PCA 
observation.  The {\it diskbb} model provides a poor fit to the artificial 
{\it XMM} data so reliable uncertainties cannot be estimated for best fit
parameters. The uncertainties (90\% confidence for one parameter) on
PCA depend on the signal-to-noise of the simulated data set, but they are of
order the symbol size or smaller for a 1.5 ksec exposure of a source at 5 kpc.
We compare the best-fit {\it diskbb} spectra to the $i=70^{\circ}$ non-LTE 
model SED's in Figure \ref{fig:fits}.

First, we focus on the EPIC-pn data sets.  The filled and open triangles mark 
the $i=45^{\circ}$ and $70^{\circ}$ measurements respectively.  Both sets
of measurements show a similar degree of spectral hardening with increasing
luminosity but with approximately the same fractional offset in $T_{\rm max}$ 
and $L_{\rm disk}/L_{\rm Edd}$ for each $l$.  The difference in apparent
luminosity between the two sets of measurements and the known $l$ values
are predominantly due to limb darkening in the atmospheres (compare
with Figure \ref{fig:limbd}), but there is also a small mismatch between our
relativistic transfer function and the $g$ correction factors we interpolate from
Zhang et al. (1997) Table 1.  
Comparing the symbols with the solid lines shows
that the derived values of $f_{\rm col}$ are inclination dependent.  When
folded through the spectral response, the
best-fit {\it diskbb} models disagree with artificial spectra by about
5-10\% at lower energy and less than 5\% near the peak.  These deviations
can be significant for the high signal-to-noise obtained in a long
exposure of the bright high/soft state X-ray binaries.  
The model provides poor fits (reduced $\chi^2 \gg 1$) to 
the data for the simulated 75 ksec exposures of sources at 5 kpc.

The PCA measurements at $l=0.1$ and 0.3 agree reasonably well with the EPIC-pn 
data sets.  The $T_{\rm max}$ values are larger because the PCA band only
extends down to 3 keV in our fits and does not include the spectral peak.
The Doppler-broadened high energy tail is consistent with a slightly higher 
inner temperature than the emission near the spectral peak which is given 
more weight in the fits to the EPIC-pn artificial spectra. The $l=0.03$ case 
is slightly more complicated because there is a spectral break above 
3 keV due the presence of bound-free opacity.  This produces a slightly
steeper fall-off in the high energy tail.  This can be more easily fit
by a lower temperature and correspondingly higher normalization leading
to a larger $L_{\rm disk}/L_{\rm Edd}$ and lower $T_{\max}$.  These shifts 
overestimate the emission near the peak which is outside of the PCA band
but must be accounted for in the EPIC-pn fits.

\section{Discussion} 
\label{discus}

\subsection{Comparison with Previous Work}

Our models are qualitatively different from those of ST95 in several important ways.
We examine the specific intensities and account for relativistic effects 
on photon geodesics while they compare the specific flux in the local frame of
the disk. Calculating specific fluxes in the local frame simplifies the 
comparison with diluted blackbodies by avoiding inclination-dependent effects.
However, this has the drawback of neglecting the non-trivial coupling 
between frequency and inclination caused by the limb darkening and
relativistic transfer which is presumably present in real accretion disks.
Another important difference between our models is our inclusion of metals.
We account for the bound-free opacity of all important ions in addition to
free-free opacity which was the only absorption opacity included in their 
calculations. The addition of metal
opacity reduces the range of $l$ in which saturated Comptonization is important
and produces modified blackbodies which are slightly softer (more consistent
with lower values of $f_{\rm col}$) than their model SED's.

ST95 conclude that multi-temperature, diluted blackbody spectra adequately
approximate their models at higher luminosities but are inadequate at
low luminosity. When they compare diluted blackbodies
with their $\alpha=0.1$, Schwarzschild black hole integrated disk spectra at
energies above 0.1 keV, the deviations are less than 20\%, 15\%, and 30\% for 
the $l=0.0057$, 0.057, and 0.57 models respectively. For the two more luminous
models, the largest disagreement is at the low energy end of the band and the 
agreement is better at higher energies.  For $l=0.0057$, the integrated
spectra show larger deviations at higher energies and the individual model
annuli are not well represented by diluted black bodies because the 
$y$-parameter is small.  They conclude that color-corrected blackbodies 
are good approximations only at the higher luminosities.

We focus on the integrated emission from the whole disk.
For all the $l=0.01$ to 0.3 models, the comparisons between diluted blackbody 
and non-LTE atmosphere model SED's depend on the viewing angle.  In section
\ref{lumin}, we presented fits to our non-LTE models with fully relativistic diluted
blackbody models with constant color correction $f_{\rm col}$.  Provided
one fits for both $f_{\rm col}$ and the normalization (in order to account
for limb darkening), these fits are surprisingly good. The disagreement
between the models is less than 20\% at most photon energies regardless
of inclination. As seen in Figure \ref{fig:lspec}, the deviations in the high
energy tails can eventually become much larger. However, by that point the 
statistical  weight of these energy bins is typically low because the 
integrated photon flux falls off rapidly in the tail.  The fitted values
of $f_{\rm col}$ at $i=70^{\circ}$ are summarized in the second column of
Table \ref{tbl1}. These best-fit values are weakly dependent on inclination,
but they are always lower than the 1.7-1.9 values reported by ST95.

A direct comparison of our models with those of Merloni et al. (2000) is 
complicated by  their assumption that a fraction $f$ of the dissipation 
occurs in the corona.
Inspection of their Figure 2 suggests that the value of $f_{\rm col}$ inferred
from their spectral models is mostly determined by the corrected disk luminosity 
parameterized by $(1-f)\dot m$. Specifically, models with different $f$ but the
same $(1-f)\dot m$ yield similar values for $f_{\rm col}$. Their $\dot m$ accounts 
for the disk efficiency so this disk luminosity parameter is the rough equivalent 
of our $l$.
Merloni et al. (2000) also construct artificial spectra from their model SED's
using an {\it RXTE} response matrix. They fit these spectra with {\it diskbb}
models and derive values of $f_{\rm col}$ by a method which is similar, but 
differs in detail
from the procedure implemented by us in section \ref{lvst} and in GD04. They 
find a trend toward increasing $f_{\rm col}$ with decreasing $(1-f)\dot m$.  If
we take the {\it diskbb} fits of section \ref{lvst} and use their method for
determining $f_{\rm col}$, we find the values listed in the fourth
column of Table \ref{tbl1}.  (For completeness, we also show values
appropriate for the {\it XMM-Newton} EPIC-pn camera in the third column.)
We find a weak increase in $f_{\rm col}$ with increasing 
$l$ which is inconsistent with their results. We suspect that this discrepancy is 
largely due to their neglect of bound-free metal opacity which causes them to
overestimate the ratio of scattering to absorption opacity.  Therefore 
$\epsilon_{\nu}$ is lower and $\tau^{\ast}_{\nu}$ is greater, providing a 
higher $T^{\ast}_{\nu}$.  The higher $T^{\ast}_{\nu}$ and lower $\epsilon_{\nu}$ 
would both give rise to harder, more modified spectra, consistent with larger 
values of $f_{\rm col}$.  These effects are most significant in the low accretion 
rate  models where Comptonization is negligible at most radii. Their
assumption of constant density likely overestimates the
density in the spectral forming regions in {\it both} gas pressured and radiation
pressure dominated annuli, increasing the ratio of absorption to
scattering opacity at $\tau^{\ast}_{\nu}$.  This reduces the $y$-parameter but
may somewhat mitigate the increased spectral hardening which comes from 
neglecting metal opacity.

\subsection{Stress Prescription and Dissipation Profile}

For the range of $l$, $\alpha$, and $a$ explored here, altering $m_0$ by
varying $\alpha$ or changing the stress prescription has little effect on
the SED.  Even changes in the vertical dissipation profile do not matter
very much unless a substantial fraction of the heating occurs above the
effective photosphere.  It therefore appears that the spectra are
insensitive to the details of the very uncertain vertical structure of
disks, which might be encouraging from a theoretical point of view, but
disappointing if one wants to use spectral observations to probe disk physics.

However, it is important to bear in mind how this robustness arises.  The
innermost parts of luminous black hole accretion disks are both radiation
pressure supported and electron scattering dominated.  Provided heat
is vertically transported by radiative diffusion, then the scale height $h$,
which determines the surface gravity $g$, is completely independent of the 
stress prescription.
If there is little dissipation in
the surface layers, the radiative flux is constant, but the gravity still
increases with height.  As a result, gradients in gas pressure are always
important in maintaining hydrostatic equilibrium near the surface, and this
always produces a very steeply declining density profile there.  This is
in contrast to the temperature profile, which is relatively 
constant in the surface layers (cf. Fig. \ref{fig:alpdepth}), and close to the 
effective temperature.
The steep decline in density occurs over a gas pressure scale height
$H_{\rm g}\sim P_{\rm gas}/(\rho g)$ (Hubeny 1990), which is independent of the
details of the overall vertical
structure as it only depends on the surface temperature and the local gravity.
The effective photosphere at characteristic frequencies is determined by
\begin{equation}
\tau^{\rm eff}=\int[3\kappa^{\rm th}(\kappa^{\rm es}+\kappa^{\rm th})]^{1/2}\rho dz=1.
\end{equation}
Because $\kappa^{\rm th}\propto\rho$ and $\kappa^{\rm es}\propto\rho^0$, this
integral is approximately proportional to
$\rho^{\ast 3/2}H_{\rm g}$.  Hence the density $\rho^\ast$ at
the effective photosphere
is approximately independent of the details of the vertical structure.  The
column mass down to the effective photosphere $m^\ast\simeq\rho^\ast H_{\rm g}$
and the temperature at the effective photosphere $T^\ast\simeq T_{\rm eff}
(\kappa^{\rm es}m^\ast)^{1/4}$ will also be approximately
independent of the vertical structure. The resulting spectrum will therefore be
quite robust.  This all fails, however,
if heat is transported by other means, such as convection,
rather than radiative diffusion.  It may also fail if substantial
density inhomogeneities are present within the disk turbulence, as we discuss
below.

The robustness of the SED's to changes in the stress is also a
feature of accretion disk models around supermassive black holes, and for
the same reason.  Figure 17 of Hubeny et al. (2000) show that accretion disk
models around a $5\times10^8$~M$_\odot$ Kerr hole are very insensitive to the
value of $\alpha$ for Eddington ratios up to at least $l\simeq0.1$.  The only
discrepancies occur above the He~II edge.  Similar differences occur in stellar
mass black hole models near the iron K edge, but this lies in the high energy
tail of the spectrum.  For high black hole spins and higher Eddington ratios,
the effective optical depth of the innermost annuli is reduced, and the effective
photosphere at all frequencies moves deeper into the disk atmosphere.  In
such cases the spectrum will start to depend on the stress and stress
prescription, with high values of $\alpha$ or an $\alpha$-disk prescription
producing harder spectra than low values of $\alpha$ or a $\beta$-disk prescription.
The dependence on $\alpha$ is illustrated for $l=0.3$ disks accreting onto
supermassive Kerr holes in Figure 16 of HBKA.  We therefore expect that high
Eddington ratio disk models around stellar mass Kerr holes to be less robust
to changes in the stress prescription as well, but we have not yet succeeded
in constructing such models.

The weak dependence of the model SED's on the choice of stress prescription
casts some doubt on GDO4's speculation that their luminosity-temperature 
relations are inconsistent with $\beta$-disk models.  Their reasoning relies on
the assumption that
Comptonization is not important in $\beta$-disks due to the larger midplane
densities inferred from a one zone model (Nannurelli \& Stella 1989). However,
the effective photospheres in our models always occur in the surface layers
where the densities are much less than the midplane density.
Therefore, Comptonization is
equally important (or unimportant depending on $l$) for both $\alpha$-disks
and $\beta$-disks over the range of parameters we consider.

In an attempt to inject first principles physics that is not present in the
standard model, we also considered modifications to the vertical dissipation
profile.  The standard model dissipation profile produces a disk which
is hydrodynamically unstable to convection in the radiation pressure dominated
regions (Bisnovatyi-Kogan \& Blinnikov 1977, HBKA),  though it is far from clear
whether significant convective transport of heat would occur in the presence
of MRI turbulence.  Magnetic field buoyancy suggests that the dissipation
may be more concentrated near the surface than in the standard model.
An annulus with a modified dissipation profile motivated by
simulations of accretion disk turbulence (Turner 2004) produces a vertical
structure that is increasingly gas pressure supported in its interior, and
hydrodynamically stable to convection.

Adopting this modified dissipation profile only produces slight changes to the
resulting spectrum, however.  The reason is that the considerations of the first
paragraph of this section still apply.  The surface density profile is still very
steep, and the physical properties of the plasma at the effective photosphere are
largely independent of the dissipation profile, provided most of that dissipation
occurs deeper than the effective photosphere.
The impact is not completely negligible as the modified annulus SED is slightly harder
than in the standard model, though the inferred $f_{\rm col}$ only differs by 
$\sim 10\%$. However, if we consider more top heavy dissipation 
profiles in which a significant fraction of the dissipation occurs near or above
the photosphere we can produce atmospheres with very different
SED's even though the surface radiative flux, gravity, and surface density 
are the same.  The ``corona''-producing annuli discussed in section \ref{corona}
are extreme examples.  These calculations may allow for a self-consistent treatment
of the hard and soft spectral components in models in which the coronal emission
is provided by flaring regions above the disk though there are several caveats to 
our treatment which remain to be addressed. These include the assumptions that the
temperatures of the ions and electrons are equivalent, the electrons are entirely 
thermal, and the corona is time-steady.

Most of the uncertainties in the vertical distribution and nature of the 
dissipation are relatively easy to address with the methods presented here. 
However, other potentially important aspects of the dynamic nature of these 
disks remain unaccounted for in our models.
Simulations (e.g. Turner 2004) suggest the presence of substantial
inhomogeneities in density due to the compressible nature of the turbulence
(Turner, Stone \& Sano 2003) and radiatively driven instabilities 
(Gammie 1998). Modeling of the emergent spectrum from such structures requires
three dimensional radiative transfer.  We have reported on Monte Carlo
radiative transfer calculations on inhomogeneous structures created by 
simulations elsewhere (Davis et al. 2003).  The temperature regime was more
appropriate for a quasar disk rather than an X-ray binary, and thermal
Comptonization was negligible. The  main effect of the inhomogeneities was to
reduce the modified blackbody effects and produce a softer, more thermal 
spectrum.  The emergent flux through an inhomogeneous medium is also larger
than the spatially averaged flux required to support the medium against
gravity (Begelman 2001).  This may reduce the vertical scale height of the
disk compared to the standard model, and make the emergent spectrum
more sensitive to the stress prescription and dissipation profile.
Because the turbulence is compressible and
supersonic, bulk Comptonization can also significantly affect the emergent
spectrum (Socrates, Davis \& Blaes 2004). In addition to the effects of Silk
damping, bulk Comptonization may also provide a radiative dissipation 
mechanism for the turbulence itself.  It is also possible that the 
time-averaged spectrum emerging from a time-dependent, turbulent annulus
differs from the spectrum emerging from a time-average of the vertical
structure.

\subsection{Observational Implications}

Our luminosity-temperature relation agrees reasonably well with the results
of GD04.  For example, between $l=0.3$ and 0.1 we find weak evolution toward 
lower $f_{\rm col}$ consistent with that observed in XTE J1550-564.  The larger
decrease in $f_{\rm col}$ in our models from $l=0.1$ to 0.03 is not clearly 
observed in their data, but it is not inconsistent due to large uncertainties 
in the fits at lower luminosities.  The magnitudes of $f_{\rm col}$ derived
using the procedure of GD04 depend on the inclination due to limb darkening.
In this case, they are in approximate agreement with the observed values
for $i=70$, which is near the measured inclination of $73.5^{+1.9}_{-2.7}$ 
for XTE J1550-564 (Orosz et al. 2002). The values of $f_{\rm col}$ are probably also
sensitive to our assumptions about inner torques, dissipation profiles, 
and metal abundances. Despite these concerns, we expect the generally weak trend 
toward  decreasing $f_{\rm col}$ with decreasing $l$ to be robust because it
is largely due to the increasing ratio of absorption to scattering opacity as the 
temperature of the disk drops.  This reduces $\epsilon_{\nu}$ and moves
$\tau^{\ast}_{\nu}$ closer to the photosphere, lowering $T^{\ast}_{\nu}$.
The spectra soften and the associated $f_{\rm col}$ values decrease.

Though we employ the concept of a color correction throughout this
paper for the purposes of comparison, we caution that
isotropic, diluted blackbodies are {\it not} perfect approximations to our 
models. 
Given the high signal-to-noise spectra afforded by an observation with 
{\it XMM-Newton}, the $\lesssim 20$\% discrepancies between the best fit, fully
relativistic diluted blackbodies and the non-LTE atmosphere spectra
are potentially significant. Any assumption of isotropy is clearly poor as 
limb darkening can alter the inferred luminosity of a source by up to 50\% 
at high inclinations. We further caution that the {\it diskbb} model is an
even poorer match to our model SED's. For high signal-to-noise observations
we do not expect it to provide adequate fits unless the disk is truncated at 
large radius. 
Even if the presence of an inner torque makes the surface brightness profile 
more consistent with the {\it diskbb} profile, the model completely neglects
relativistic transfer effects.  The best-fit 
{\it diskbb} spectra are always significantly narrower than our models due to 
differences in the surface brightness profiles and relativistic broadening.
To  accurately model all the emission from a disk component, one 
needs to include the relativistic effects on the disk structure and transfer 
function in the spectral model {\it before} fitting it to the observed SED.
The data in Table \ref{tbl1} indicate 
that different but seemingly consistent methods of calculating $f_{\rm col}$
with {\it diskbb} model fits give differing results. Furthermore, the results
depend on the detector response and band.

These concerns are particularly relevant for higher signal-to-noise 
observations which 
have inferred relativistically broadened Fe K$\alpha$ lines but fit the disk 
component with an unbroadened {\it diskbb} model (Miller et al. 2004a, 
2004b).
In some cases the iron line profiles are so broad and asymmetric that they 
require a near maximally spinning black hole with a very steep iron line radial 
emissivity profile. Contrary to our expectations of 10-15\% discrepancies, the
{\it diskbb} model seems to account surprisingly well for the presumed disk 
component in these observations.  We suspect this may be due in part to 
their use of a power law to model 
the coronal emission.  If the disk provides the seed photons for the corona, a
power law will likely overestimate the flux of this component at the low 
energy end of the band where we predict the discrepancies are largest.  This 
extra emission could account for the difference between our model SED's and 
the {\it diskbb} models.  We test this hypothesis by constructing an artificial 
spectrum with our $l=0.1$, $a=0.998$, and $i=45^{\circ}$ model SED, assuming it
is located at 5 kpc and accounting for an intervening H column of 
$5 \times 10^{21} \, \rm cm^{-2}$. We also include a {\it compTT} component 
whose flux is about 10\% of the model flux.  First, we fit these data with a
model that includes a {\it diskbb} component, a {\it compTT} component with 
the seed temperature fixed at the input value, and neutral absorption at the
input H column.  We get a poor fit with 
residuals of $\sim 15$\%. When we replace the {\it compTT} model with a 
power law, the residuals of the best fit are reduced to less than 5\% and the 
quality of fit improves dramatically. It therefore seems plausible that the
combinations of a {\it diskbb} and power law may effectively reproduce the
SED of a real disk even though the {\it diskbb} model does not properly
account for the accretion disk emission.

\section{Conclusions}
\label{conc}

We have calculated several representative non-LTE, fully relativistic
models for geometrically thin disks accreting onto a $10 M_{\odot}$ black hole.
We first explored the behavior of standard model (SS73, NT73) accretion disks.
We found that the inclusion of abundant metals has
a significant impact on the model SED's because of bound-free absorption opacity
at high photon energies.  The inclusion of this additional opacity generally
decreases the importance of Comptonization, but still produces softer, more
Wien-like spectra than would be expected from free-free opacity alone. The
associated diluted blackbody color corrections are generally lower than found
by previous authors (ST95, Merloni et al. 2000).

The frequency dependence of the disk integrated model SED's is qualitatively
similar to that of spectra produced by assuming isotropic diluted blackbodies
with constant color correction for the local emission, provided these local
spectra are folded through a relativistic transfer function.  Quantitative
differences still exist which should be discernible for some observations with
modern X-ray observatories.  The effects of limb darkening are
significant and the assumption of isotropy of the local emission must be
compensated by fitting the normalization as well as the color correction.
The best-fit $f_{\rm col}$'s are weakly dependent on the 
black hole spin and accretion rate, generally increasing as either
parameter increases.

We have found that the standard model SED's are strongly affected by changes
in accretion rate and black hole spin, but they are only weakly dependent on the
value of $\alpha$ for the parameter space explored here.  Modifying the
stress prescription by replacing $P$ with $P_{\rm gas}$ in equation
\ref{eq:stress} also has little effect on the model SED's, contrary to the
supposition of GD04.  The reason for this is that the overall disk scale
height in the inner, radiation pressure supported regions is largely
independent of the stress prescription, and so are the resulting physical
conditions at the effective photosphere.  Higher Eddington ratio models,
particularly those around maximally spinning holes, may be more sensitive to
the treatment of the stress.  Steep density gradients generally exist in
the surface layers, and modeling the disk annuli with constant density
slabs is also a poor approximation.

We found that our model SED's are generally not well matched by a simple
{\it diskbb} model over the whole 0.3-10 keV band.  Fits using the
{\it XMM-Newton} EPIC-pn response have discrepancies of greater than 10\%.
Such large discrepancies are generally not seen in {\it diskbb} fits to high
signal-to-noise observations of X-ray binaries with relativistically broadened
Fe K$\alpha$ lines.  We speculate that the lack of residuals in these fits may
be due to overestimation of the coronal emission at low energies. A resolution
of this issue will require detailed fits of our models to real observations.

We used the method of GD04 to construct a luminosity-temperature relation
for our Schwarzschild disk models.  The luminosity scales as the fourth
power of the temperature, but we also found a trend toward larger
$f_{\rm col}$ with increasing $l$ which is generally consistent with
their analysis for several sources.  The agreement with their measured
values of $f_{\rm col}$ is somewhat more uncertain and our $f_{\rm col}$
measurements depend on the inclination due to limb darkening effects.
The values of $f_{\rm col}$ which we infer using GD04's method are higher
than those we derive from fits with the relativistic diluted blackbody
models (see Table 1).

We have also looked at modifications of the standard model based on recent
theoretical progress in understanding magnetohydrodynamic stresses in the
flow.  Adopting a vertical dissipation profile derived from numerical
simulations produces a large change in the vertical disk structure but 
has a weaker effect on the 
SED, consistent with a $\sim 10\%$ increase in the best-fit $f_{\rm col}$.
Profiles which have a greater fraction of the dissipation near the
surface can produce much larger changes. Thus, our choice of dissipation
profiles represents a large uncertainty in the models.

The addition of an 
inner torque at fixed fraction of the Eddington luminosity produces 
a significant hardening of the spectrum similar to that made by increasing the 
black hole spin.  Unless the two types of models are distinguished by fits 
with detailed continuum models or relativistic line profiles, it may be 
difficult to infer either parameter without making an assumption about the 
other.  Additional uncertainties in the disk physics that we have not explored here
are the possibility of bulk vertical transport of heat, e.g. convection,
and the effects of density inhomogeneities and bulk Comptonization
in the supersonic disk turbulence.

\acknowledgements{We thank Phil Chang, Chris Done, and Aristotle Socrates 
for very useful discussions and suggestions. This work was supported by 
NASA grant NAG5-13228.}

\clearpage

\begin{deluxetable}{lccccc}
\tablecolumns{6}
\tablecaption{Color Corrections \label{tbl1}}
\tablewidth{0pt}
\tablehead{
\colhead{$l$} &
\colhead{$f_{\rm FR}$} &
\colhead{$f_{\rm MFR}^{\rm X}$} &
\colhead{$f_{\rm MFR}^{\rm R}$} &
\colhead{$f_{\rm GD}^{\rm X}$} &
\colhead{$f_{\rm GD}^{\rm R}$}
}

\startdata
0.3 &
1.64 &
1.62 &
1.67 &
1.88 &
1.97 \\
0.1 &
1.56 &
1.60 &
1.63 &
1.85 &
1.90 \\
0.03 &
1.46 &
1.50 &
1.36 &
1.74 &
1.58 \\
0.01 &
1.40 &
1.29 &
\nodata &
1.50 &
\nodata \\

\enddata

\tablecomments{Calculations of $f_{\rm col}$ based on fits to our model spectra
for $\alpha=0.1$ disks around Schwarzschild black holes, viewed at an
inclination of $70^\circ$.  Each column uses a different method to determine
$f_{\rm col}$ (see text).  The second column summarizes $f_{\rm col}$ from our fits to 
the non-LTE model SED's with our fully relativistic diluted blackbody 
model spectra.  A superscript X or R indicates the $f_{\rm col}$ values were
calculated using the {\it diskbb} fits to the {\it XMM-Newton} EPIC-pn or 
{\it RXTE} PCA data sets respectively. A subscript MFR or GD04 means that we 
use the methods described in Merloni et al. (2000) or GD04 (respectively)
to calculate $f_{\rm col}$.}

\end{deluxetable}

\clearpage

\begin{thebibliography}{}

\bibitem[Abramowicz et al. 1998]{abr98} Abramowicz, M. A., Czerny, B.,
	Lasota, J. P., \& Szuszkiewicz, E. 1988, \apj, 332, 646

\bibitem[Agol 1997]{ago97} Agol, E. 1997, Ph.D. thesis, Univ. California,
Santa Barbara

\bibitem[Agol \& Krolik 1998]{aak98} Agol, E., \& Krolik, J. 1998, \apj, 507, 304

\bibitem[Agol \& Krolik 2000]{aak00} Agol, E., \& Krolik, J. H. 2000, \apj, 528, 161

\bibitem[Balbus \& Hawley 1991]{bah91} Balbus, S. A. \& Hawley, J. F. 1991, 
        \apj, 376, 214

\bibitem[Begelman 2001]{beg01} Begelman, M. C. 2001, \apj, 551, 897

\bibitem[Bisnovatyi-Kogan \& Blinnikov 1977]{bis77} Bisnovatyi-Kogan, G. S., \&
Blinnikov, S. I. 1977, A{\&}A, 59, 111

\bibitem[Chandrasekhar 1960]{cha60} Chandrasekhar, S. 1960, Radiative Transfer
	(New York: Dover)

\bibitem[Cunningham 1975]{cun75} Cunningham, C. T. 1975, \apj, 202, 788

\bibitem[Cunningham 1976]{cun76} Cunningham, C. T. 1976, \apj, 208, 534

\bibitem[Davis et al. 2004]{dav04} Davis, S. W., Blaes, O. M., Tuner, N. J.,
        \& Socrates, A. 2004, in ASP Conf. Ser. 311, AGN Physics with the
	Sloan Digital Sky Survey, ed. G. T. Richards \& P. B. 
	Hall(San Francisco: ASP), 135

\bibitem[Gammie 1998]{gam98} Gammie, C. F. 1998, MNRAS, 297, 929

\bibitem[Gammie 1999]{gam99} Gammie, C. F. 1999, ApJ, 522, L57

\bibitem[Gierli\'nski \& Done 2004]{gad04} Gierli\'nski, M., \& Done, C. 2004,
	\mnras, 347, 885 (GD04)

\bibitem[Gierli\'nski et al. 1999]{gie99} Gierli\'nski, M., Zdziarski, A. A.,
	Poutanen, J., Coppi, P.S., Ebisawa, K., \& Johnson W.N. 1999, \mnras,
	309, 496	

\bibitem[Hawley \& Krolik 2002]{haw02} Hawley, J. F., \& Krolik, J. H. 2002,
ApJ, 566, 164

\bibitem[Hubeny 1990]{hub90} Hubeny, I. 1990, \apj, 351, 632

\bibitem[Hubeny et al. 2000]{hub00} Hubeny, I., Agol, E., Blaes, O., \&
	Krolik, J.H. 2000 \apj, 533, 710

\bibitem[Hubeny et al. 2001]{hub01} Hubeny, I., Blaes, O., Krolik, J.H., \&
        Agol, E. 2001 \apj, 559, 680 (HBKA)

\bibitem[Hubeny \& Hubeny 1997]{hah97} Hubeny, I., \& Hubeny, V. 1997
	 \apj, 484, L37

\bibitem[Hubeny \& Hubeny 1998]{hah98} Hubeny, I., \& Hubeny, V. 1998
	 \apj, 505, 558 (HH98)

\bibitem[Hubeny \& Lanz 1995]{hal95} Hubeny, I., \& Lanz, T. 1995, \apj,
	439, 875

\bibitem[Krolik 1999]{kro99} Krolik, J. H. 1999, ApJ, 515, L73

\bibitem[Lightman \& Eardley 1974]{lae74} Lightman, P., \& Eardley, D. M.
	1974, \apj, 187, L1

\bibitem[Merloni et al. 2000]{mer00} Merloni, A., Fabian, A.C. \& Ross, R.R.
	2000, \mnras, 313, 193

\bibitem[Miller et al. 2004a]{mi04a} Miller, J. M., et al. 2004a, \apj, 601,
	450

\bibitem[Miller et al. 2004b]{mi04b} Miller, J. M., et al. 2004b, \apj, 606,
	L131

\bibitem[Mitsuda et al. 1984]{mit84} Mitsuda, K., et al. 1984, \pasj, 36, 741

\bibitem[Nannurelli \& Stella 1989]{nas89} Nannurelli, M., \& Stella, L. 1989,
	\aap, 226, 343

\bibitem[Novikov \& Thorne 1973]{nat73} Novikov, I.D. \& Thorne, K.S. 1973, 
	in Black Holes, eds. C. De Witt and B. De Witt (New York: Gordon 
	\& Breach) p. 343 (NT73)

\bibitem[Orosz et al 2002]{oro02} Orosz J. A. et al. 2002, \apj, 568, 845
\bibitem[Paczy\'nski \& Witta 1980]{paw80} Paczy\'nski B., \& Witta P. J.
	1980 , \aap, 88, 23

\bibitem[Page \& Thorne 1974]{pat74} Page, D.N., \& Thorne, K.S. 1974, \apj, 
 	191, 499 

\bibitem[Riffert \& Herold 1985]{rah85} Riffert, H. \& Herold, H. 1995, \apj, 
	450, 508

\bibitem[Ross \& Fabian 1993]{raf93} Ross, R. R. \& Fabian, A. C. 1993, \mnras,
	261, 74

\bibitem[Sakimoto and Coroniti 1981]{sac81} Sakimoto, P.J., \& Coroniti, 
	F.V. 1981, ApJ, 247, 19

\bibitem[Shakura \& Sunyaev 1973]{sha73} Shakura, N. I., \& Sunyaev, R. A.
1973, \aap, 24, 337 (SS73)

\bibitem[Shimura \& Takahara 1995]{sat95} Shimura, T., \& Takahara, F. 1995,
	\apj, 445, 780 (ST95)

\bibitem[Socrates, Davis \& Blaes 2004]{soc04} Socrates, A., Davis, S. W.,
\& Blaes, O. 2004, ApJ, 601, 405

\bibitem[Stella \& Rosner 1984]{sar84} Stella, L., \& Rosner, R. 1984, \apj, 
	277, 312  

\bibitem[Svensson \& Zdziarski 1994]{saz94} Svensson, R., \& Zdziarski, A. A.
	1994, \apj, 436, 599

\bibitem[Thorne 1974]{tho74} Thorne, K. S. 1974, \apj, 191, 507

\bibitem[Titarchuk 1994]{tit94} Titarchuk, L. 1994, \apj, 434, 570

\bibitem[Turner 2004]{tur04} Turner, N. J. 2004, \apjl, 605, L45

\bibitem[Turner et al. 2003]{tur03} Turner, N. J., Stone, J. M., Krolik,
	J. H., \& Sano, T.  2003, \apj, 593, 992

\bibitem[Wang et al. 1999]{wan99} Wang, J. M., Szuszkiewicz, E., Lu, F. J., \&
	Zhou, Y. Y. 1999, \apj, 522, 839

\bibitem[Zhang, Cui, \& Chen 1997]{zcc97} Zhang, S. N., Cui, W., \& Chen, W.
	1997, \apjl, 482, L155

\end{thebibliography}
\end{document}